\documentclass[fleqn,usenatbib]{mnras}

\usepackage{newtxtext,newtxmath}

\usepackage[T1]{fontenc}
\usepackage{ae,aecompl}

\usepackage{graphicx}	
\usepackage{amsmath}	
\usepackage{epsfig}
\usepackage{url}
\usepackage{widetext}
\usepackage[normalem]{ulem}

\newcommand{\rev}[1]{\textbf{#1}}

\title[Variation in jet PA from jet precession in OJ~287]{Explaining temporal variations in the jet position angle of the blazar OJ~287 using its binary black hole central engine model}

\author[Dey et al.]{
Lankeswar Dey,$^{1}$\thanks{Email: lankeswar.dey@tifr.res.in}
Mauri J. Valtonen,$^{2,3}$
A. Gopakumar,$^{1}$
Rocco Lico,$^{4,5}$
Jos{\'e} L. G{\'o}mez,$^{4}$
\newauthor{}
Abhimanyu Susobhanan,$^{1}$
S. Komossa,$^{5}$
and Pauli Pihajoki$^{6}$
\\
$^{1}$Department of Astronomy and Astrophysics, Tata Institute of Fundamental Research, Mumbai 400005, India\\
$^{2}$Finnish Centre for Astronomy with ESO, University of Turku, Finland\\
$^{3}$Department of Physics and Astronomy, University of Turku, Turku, Finland\\
$^{4}$Instituto de Astrof\'{\i}sica de Andaluc\'{\i}a-CSIC, Glorieta de la Astronom\'{\i}a s/n, 18008 Granada, Spain\\
$^{5}$ Max-Planck-Institut f\"ur Radioastronomie, Auf dem H\"ugel 69, 53121 Bonn, Germany\\
$^{6}$Department of Physics, University of Helsinki, Gustaf H{\"a}llstr{\"o}min katu 2a, FI-00560, Helsinki, Finland
}

\date{Accepted XXX. Received YYY; in original form ZZZ}

\begin{document}
\label{firstpage}
\pagerange{\pageref{firstpage}--\pageref{lastpage}}
\maketitle

\begin{abstract}
The bright blazar OJ~287 is the best-known candidate for hosting a supermassive black hole binary system. It inspirals due to the emission of nanohertz gravitational waves (GWs). 
Observations of historical and predicted quasi-periodic high-brightness flares in its century-long optical lightcurve, allow us to determine the orbital parameters associated with the binary black hole (BBH) central engine.
In contrast, the radio jet of OJ~287 has been covered with
Very Long Baseline Interferometry (VLBI) observations for only about $30$ years and these observations reveal that the position angle (PA) of the jet exhibits temporal variations at both \textit{millimetre} and \textit{centimetre} wavelengths. 
Here we associate the observed PA variations in OJ~287 with the precession of its radio jet. 
In our model, the evolution of the jet direction can be associated either with the primary black hole (BH) spin evolution or with the precession of the angular momentum direction of the inner region of the accretion disc.
Our Bayesian analysis shows that the BBH central engine model, primarily developed from optical observations, can also broadly explain the observed temporal variations in the radio jet of OJ~287 at frequencies of 86, 43, and 15 GHz.
Ongoing Global mm-VLBI Array (GMVA) observations of OJ~287 have the potential to verify our predictions for the evolution of its $86$ GHz PA values. 
Additionally, thanks to the extremely high angular resolution that the Event Horizon Telescope (EHT) can provide, we explore the possibility to test our BBH model through the detection of the jet in the secondary black hole .
\end{abstract}

\begin{keywords}
Blazar: individual (OJ~287) -- accretion disc -- jets -- black hole physics -- binaries (general) 
\end{keywords}



\section{Introduction}
\label{sec:intro}

Active galactic nuclei (AGNs) are the most energetic persistent sources known to mankind \citep{KN99}.
They are the luminous central regions of certain galaxies, known as active galaxies, in which a supermassive black hole (SMBH) accretes matter from the surrounding accretion disc, ensuring that enormous amounts of energy are emitted from a very small region \citep{lyn69}. 
Some AGNs launch jets, which are collimated beams of charged particles, accelerated to relativistic velocities.
Depending on the angle between the jet and our line of sight, AGNs appear differently \citep{Urry95}.
AGNs viewed along a line of sight very close to the jet axis, known as blazars, are dominated by the radiation from the jet, making it very difficult to observe the host galaxy \citep{GG98}.
The launching mechanism of jets associated with AGNs has remained one of the unsolved problems in modern astrophysics. 
Though different models have been proposed to explain the jet launching mechanism \citep[e.g.,][]{BZ77,BP82}, its details have remained elusive.

OJ~287 is a bright unique blazar situated at a redshift of $z = 0.306$. 
Its optical observations date back to the 1880s \citep{sil1988}, and the extended optical lightcurve spanning 130 years shows intriguing quasi-periodic magnitude variations.
 In its apparent magnitude, there exists a long-term periodic variation with a period of $\sim$60 years \citep{val06b}. 
Additionally, the light curve displays quasi-periodic doubly-peaked high-brightness flares with a period of $\sim$12 years \citep{val06b,dey18}. 
These unique magnitude variations in the optical light curve of OJ~287 can  be explained with the help of our binary black hole (BBH) central engine model \citep{LV96,sun97,dey19a}. 
According to this model, a supermassive secondary black hole (BH) is orbiting around a much more massive primary BH in a precessing eccentric orbit with a redshifted orbital period of $\sim$12 years \citep{dey18}. 
The orbital plane is assumed to be at an angle to the accretion disc of the primary BH, and the flares happen when the secondary BH collides with the disc. 
For the last 20 years, this model has been very successful in predicting the impact flares in the optical light curve of this unique blazar  \citep{val07,val11a,dey18}.
Specifically, the BBH central engine model has successfully predicted the observed impact flares of 2007, 2015 and 2019 \citep{val08,val16,laine20}.

In contrast to optical observations, the jet of OJ~287 has been under regular observations in high-frequency radio waves for only the last $30$ years \citep{agu12,coh17,hodgson17,tat13}. 
At lower frequencies (8 GHz and 5 GHz), sparse observations of OJ~287 were done in earlier epochs \citep{roberts87, gabuzda89, tateyama99}.
Generally, blazars display high variability in radio wavelengths, and OJ~287 is no exception.
However, the position angle (PA) of the projected jet of OJ~287 on the sky plane shows certain systematic variations and sudden jumps \citep{agu12, coh17}. 
Such variations, observed at different radio frequencies, also turned out to be correlated.
The overall trend for the last three decades is that the PA decreases with time. 
However, OJ~287's PA experienced a rapid jump by $\sim 130^\circ$ during $2004$ at $43$ GHz \citep{agu12}. A similar kind of jump was also observed at $15$ GHz in $2010$ \citep{coh17}.

A precessing jet provides a natural explanation for the observed temporal variations in the PAs of radio jets in quasars \citep{abraham00, caproni04, tateyama04, bri18, qian18}.
In the case of
OJ~287, \citet{agu12} observed a gradual rise and then a sharp fall in the 43 GHz core flux density during its PA jump at 43 GHz during 2004.
This also points towards a precessing jet where the jump in PA happens when the precessing jet makes a close approach to the line of sight.
Further, a small change in OJ~287's jet direction can lead to a significantly larger change in its radio jet's PA.
This is due to the small angle between the jet of OJ~287 and our line of sight, characteristic of blazars.
Most previous studies have tried to model the high-frequency radio observations of OJ~287 while ignoring the detailed description of the system, developed from its long-term optical observations.
Naturally, these models tend to be fairly unsuccessful in describing the behaviour of OJ~287 at optical wavelengths.
In this paper, we explore the possibility of explaining the observed PA variations of OJ~287's radio jet while employing the BBH central engine model, mainly developed for describing the major variations in its optical lightcurve \citep{LV96, val08, val11a, dey18}.
The present study aims to model simultaneously the observed PA variations at three different radio frequencies, namely 86 GHz, 43 GHz, and 15 GHz, for the first time. 
Note that observations at different radio frequencies probe the nature of the jet at different distances from the jet base. Higher frequencies probe a region closer to the base, compared to lower frequencies \citep{pushkarev2012}.

We model the jet precession in OJ~287 by invoking the following two alternative scenarios. 
In our first approach (the \textit{spin model}), we let the spin evolution of the primary BH determine the radio jet direction \citep{BZ77}.
In our BBH central engine model for OJ~287, the spin angular momentum of the primary BH and the BBH orbital angular momentum are not aligned.
This forces the primary BH spin to precess about the direction of the total angular momentum vector mainly due to general relativistic spin-orbit interactions \citep{KG05,val10}. 
We employ post-Newtonian (PN) accurate equations to evolve simultaneously the BBH orbit and the primary BH spin while linking the evolution of radio jet direction with that of the primary BH spin in our \textit{spin model} \citep{bla14, dey18}.

For the second scenario (the \textit{disc model}), we model the temporal evolution of the angular momentum of the inner region of the accretion disc and link it to the radio jet's PA variations \citep{BP82}.
We model the accretion disc as a cloud of point particles  interacting via a grid-based viscous force \citep{pih13}.
Additionally, these disc particles follow PN accurate equations of motion and all of them interact gravitationally with both BHs. 
Such a prescription allows us to follow the evolution of the angular momentum of the inner region of the accretion disc in a computationally efficient and physically realistic manner.
In this scenario, the direction of the jet is assumed to be determined by the angular momentum of the inner region of the accretion disc.

We pursue these two distinct approaches as the jet launching mechanisms in AGNs and what determines the observed jet evolution are not understood in great detail.
In both scenarios, we employ a Bayesian framework to estimate the model parameters used to interpret the variations of OJ 287’s jet PA at three different radio frequencies.
We note in passing that the second scenario and its implications were probed in an earlier investigation \citep{VW12,VP13}.
The present investigation improves their descriptions  by introducing the effects of viscosity in the accretion disc and goes on to model the influence of primary BH spin precession on the accretion disc. 
Additionally, it incorporates newer PA datasets on OJ~287's jet, observed at three different radio wavelengths in contrast to the previous studies.
In particular, the inclusion of the 86 GHz data for the first time, to model the jet precession in OJ~287 is a major advancement in our analysis as it tracks the evolution of the innermost part of the radio jet in this blazar.

Interestingly, OJ~287 is a potential non-horizon target for the Event Horizon Telescope (EHT), which recently published the first image of the supermassive black hole present at the centre of M87 \citep{EHT19a}. 
The $\sim$20$\,\mu$as nominal resolution of the EHT at 230~GHz is at the limit of the required resolution to resolve the binary black hole system in OJ~287 at its maximum orbital separation on the plane of the sky.
However, we argue that the on-going and future EHT campaigns to observe OJ~287 may have the potential to substantiate the BBH scenario for OJ~287. 
This tentative claim requires that the secondary BH in OJ~287 support a temporary jet, as noted by \citet{pih13} and we explore this idea in detail in this paper.

We structure the paper in the following way.  Section~\ref{sec:pa_data} specifies various PA datasets that we employ.
Section~\ref{sec:bhh_spin} provides a summary of our BBH model for OJ~287 and how we model the spin precession of the primary BH. 
Our modelling of the accretion disc and its evolution are  described in Section~\ref{sec:disc_model}. 
How we make contact with  the available radio jet observations \rev{is} detailed in Section~\ref{sec:obs_implication}. 
This section also explores the tentative implications of our BBH scenario for the ongoing and upcoming EHT campaigns on OJ~287.
We summarise our efforts and discuss their implications in Section~\ref{sec:discussion}.

\section{PA datasets at different radio frequencies}
\label{sec:pa_data}

To model the jet precession inherent in the binary central engine scenario for OJ~287, we employ the jet PA of OJ~287 observed at three different frequencies: 15, 43 and 86\,GHz.
At 15 and 43 GHz, we use PA measurements provided by \citet{coh17} and \citet{hodgson17} spanning a time range from 1995 to 2015.
The error bars on the PA values for 15 and 43 GHz are not available in \citet{coh17} and we used a fixed error bar of $5^{\circ}$ for all 15 and 43 GHz data points following \citet{hodgson17}.
At 86 GHz, we use the Global Millimetre VLBI Array (GMVA) images for a total of 11 epochs extending from 2008 to 2017. 
To constrain the jet PA, we perform the ridgeline analysis  following the approach presented in \citet{Pushkarev2017} and more recently in \citet{Lico2020}, using the \texttt{HEADACHE}\footnote{Available at \url{https://github.com/junliu/headache}} python package developed by Jun Liu. 
After convolving each GMVA image with a common circular beam of 0.1 milli-arcseconds (mas) radius, we take slices across the jet in steps of 0.01 mas along the jet direction, and we look for the maximum in the flux density.
By averaging all the ridgeline points we determine the final PA value and the standard deviation is taken to be the 1$\sigma$ uncertainty.
We confirm that for the 7 observing epochs between 2008 and 2012, the jet PA values obtained by using our method are consistent with those presented in \citet{hodgson17}.

\section{Jet precession inherent in our BBH central engine description for OJ~287}

The BBH central engine model, as noted earlier, provides a natural explanation for the observed outbursts from OJ~287 in the optical wavelengths.
In this model, some of these synchrotron flares arise due to tidally induced mass flows in the primary BH's accretion disc which are caused by the pericentre passage of the secondary BH and occur at $\sim 12$ year intervals \citep{sun97}.
Additionally, its quasi-periodic doubly peaked optical flares with a lifetime of a month or so are superposed upon the tidal flares.
These doubly-peaked flares arise when the secondary BH impacts the accretion disc of the primary BH twice every orbit \citep{LV96}, generating hot gas bubbles that emerge on both sides of the accretion disc.
The resulting hot gas bubbles expand and cool with time and radiate strongly after becoming optically thin. 
The emerging radiation, observed as flares from OJ~287, rises sharply in the course of a few hours and is mainly produced by thermal bremsstrahlung \citep{LV96,dey19a}. 
Furthermore, the time delay between the disc crossing and the appearance of such flares depends on various properties of the BBH central engine \citep{LV96}.
In our model, these outbursts provide certain fixed points of the BBH orbit which allows us to extract various parameters of the BBH central engine \citep{val07,dey18,val19}.
Observations of three predicted impact flares during the last 15 years allowed us to constrain many ingredients of the BBH central engine including the accretion disc parameters \citep{val19}.
These considerations prompted us to employ our BBH central engine model for OJ~287 to explain the multi-epoch high-frequency radio observations of its jet. 
In what follows, we provide two alternative descriptions for the temporal evolution of the radio jet in OJ~287 and connect them with observations.

\subsection{Spin model for OJ~287's jet}
\label{sec:bhh_spin}
This model, as noted earlier, assumes that the radio jet of OJ~287 is aligned with the direction of the primary BH spin of our binary system.
Clearly, we need to model accurately the precession of the primary BH spin, and the relevant expression for the precession of the primary BH spin vector ${\mathbf{s}}_1$ reads
\begin{eqnarray}
\frac{d{\mathbf{s}}_1}{dt} &=& 
\left ( {\mathbf{\Omega}}_{\rm SO} + {\mathbf{\Omega}}_{\rm Q} \right ) 
 \times {\mathbf{s}}_1 
\,,
\label{eqn:spinprecession}
\end{eqnarray}
\noindent
where $\mathbf{s}_1$ is the direction of of the primary BH spin angular momentum, and ${\mathbf{\Omega}}_{\rm SO}$ and ${\mathbf{\Omega}}_{\rm Q}$ represent relativistic and classical spin-orbit interactions  \citep{BC79,val10,val11b}.
The spin angular momentum of the primary BH is given by  ${\mathbf{S}}_1 = G\, m_1^2 \, \chi_1 \, {\mathbf{s}}_1/c$, where $m_1$ and $\chi_1$ are the mass and the Kerr parameter of the primary BH (in general relativity, $\chi_1$ can have values between $0$ and $1$).
The form of equation~(\ref{eqn:spinprecession}) ensures that the magnitude of ${\mathbf{S}}_1$ remains a constant while its direction given by the unit vector ${\mathbf{s}}_1$ experiences precession.
The temporal evolution of various dynamical variables that appear in the above precession equation for ${\mathbf{s}}_1$ is provided by the PN accurate orbital dynamics of the binary.
Recall that the PN approximation to general relativity provides corrections to the leading order Newtonian orbital dynamics in terms of a small parameter ${(v/c)}^2$, where $v$ is the typical orbital velocity and $c$ is the speed of light \citep{bla14,WM17}. 
In the centre of mass frame, the equations of motion  can be schematically written as 
\begin{eqnarray}
    \mathbf{a} \equiv \frac{d^2 \mathbf{x}}{dt^2} &=& \ddot{\mathbf{x}}_{0} + \ddot{\mathbf{x}}_{\rm 1PN} + \ddot{\mathbf{x}}_{\rm 2PN} + \ddot{\mathbf{x}}_{\rm 3PN} \nonumber\\
    &&+ \ddot{\mathbf{x}}_{\rm 2.5PN} + \ddot{\mathbf{x}}_{\rm 3.5PN} + \ddot{\mathbf{x}}_{\rm 4PN(tail)} + \ddot{\mathbf{x}}_{\rm 4.5PN} \nonumber\\
    &&+ \ddot{\mathbf{x}}_{\rm SO} + \ddot{\mathbf{x}}_{\rm Q}\ ,
    \label{eqn:eom}
\end{eqnarray}
where $\mathbf{x}$ is the centre of mass relative separation vector between the two BHs and the $\ddot{\mathbf{x}}_{0}$ term represents the familiar Newtonian inverse-square acceleration.
In the above equation, the first line represents the conservative general relativistic contributions to the orbital dynamics that causes the advance of the pericentre. 
The corrections in the second line stand for the effects of GW emission on the orbital dynamics while the general relativistic and classical spin-orbit contributions are denoted by 
$\ddot{\mathbf{x}}_{\rm SO}$ and $\ddot{\mathbf{x}}_{\rm Q}$, respectively.
Detailed discussions about these PN contributions are provided in \citet{bla14}, \citet{WM17}, and \citet{dey18}.
We note in passing that such a general relativistic description for the BBH orbit is also employed to track the secondary BH trajectory to explain the optical wavelength observations of OJ~287.
For the present investigation, we let $m_1= 18.3 \times 10^9 M_{\odot}$, $m_2= 150 \times 10^6 M_{\odot}$, $P_{\rm orb}= 12.06$ {years},  $e=0.657$, and $\chi_1= 0.38$, as estimated in \citet{dey18}, while tracking the evolution of 
${\mathbf{s}}_1$.

\begin{figure}
    \centering
    \includegraphics[width=0.9\columnwidth]{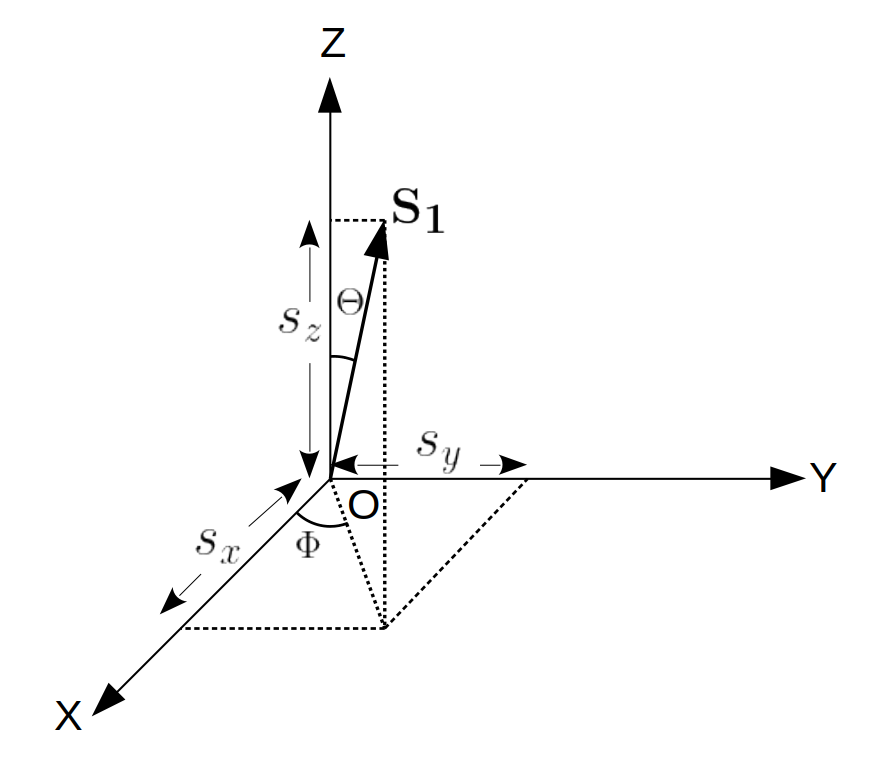}
    \caption{The coordinate system used to calculate the evolution of  the primary BH spin direction in OJ~287. The accretion disc lies in the $XY$ plane and $\mathbf{S_1}$ is the spin angular momentum of the primary BH. The angles $\Theta$ and $\Phi$ specify  the direction of the primary BH spin which are denoted by $\Theta_{\rm spin}$ and $\Phi_{\rm spin}$ in the text.
    In the case of the \textit{disc model}, $\mathbf{S_1}$ represents the average angular momentum of the inner part of the accretion disc and the angles $\Theta$ and $\Phi$ are then referred to as $\Theta_{\rm disc}$ and $\Phi_{\rm disc}$, respectively.}
    \label{fig:coordinate}
\end{figure}

\begin{figure}
	\includegraphics[width=\columnwidth]{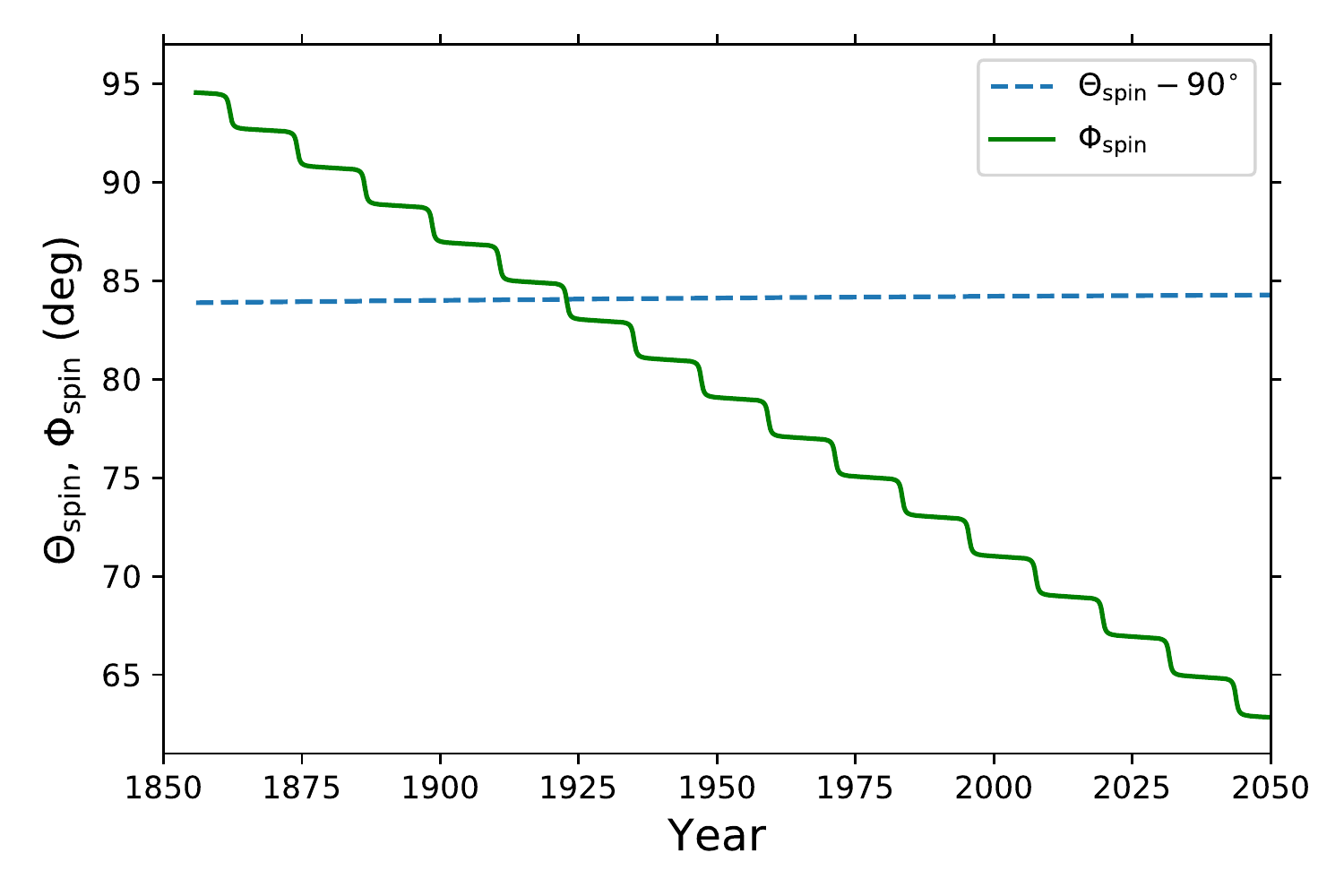}
    \caption{Temporal evolution of the primary BH spin in OJ~287 binary, specified by its polar and azimuthal angles $\Theta_{\rm spin}$ and $\Phi_{\rm spin}$. For the $\Theta_{\rm spin}$ variations, we plot the evolution of $\Theta_{\rm spin} - 90^{\circ}$ in time. The angle $\Phi_{\rm spin}$ changes with time but $\Theta_{\rm spin}$ appears to remain a constant.    }
    \label{fig:spin_prec}
\end{figure}

We compute the evolution of the primary BH spin by employing the above listed general relativity-based equations for ${\mathbf{s}}_1$ for a time window spanning around 600 years.
This is done by monitoring the temporal evolution of certain polar and azimuthal angles, $\Theta_{\rm spin}$ and $\Phi_{\rm spin}$, that specify the direction of the spin angular momentum of the primary BH in a reference frame (Figure~\ref{fig:coordinate}). 
This reference frame is defined in such a way that the accretion disc lies on the $XY$ plane and we track the temporal evolution of all three Cartesian components of the spin angular momentum ($s_x$, $s_y$, and $s_z$ in Figure~\ref{fig:coordinate}) in such a frame.
The angle between the primary BH spin angular momentum \rev{($\mathbf{S}_1$)} and the $Z$ axis is defined as $\Theta_{\rm spin}$ while $\Phi_{\rm spin}$ defines the angle between the $X$ axis and the projected spin direction on the $XY$ plane.
This allows us to extract straightforwardly the evolution of $\Theta_{\rm spin}$ and $\Phi_{\rm spin}$ as a function of time, as plotted in Figure~\ref{fig:spin_prec}.
We clearly see non-uniform temporal evolution for the azimuthal angle $\Phi_{\rm spin}$ of the primary BH spin over a period of 200 years between 1850 and 2050 in Figure~\ref{fig:spin_prec}.
However, the polar angle $\Theta_{\rm spin}$ appears to be a constant during the above time window.
This is because the primary BH spin precesses around the total angular momentum of the binary system \citep{KG05}, and the $Z$ axis happens to lie very close to the total angular momentum direction. 
Although $\Theta_{\rm spin}$ does not vary noticeably over the time span plotted in Figure~\ref{fig:spin_prec}, it varies over a much longer timescale of $\sim$1000 years.
Note that we plot $\Theta_{\rm spin} - 90^{\circ}$ to display the  temporal evolution for both $\Phi_{\rm spin}$ and $\Theta_{\rm spin}$ in the same plot.
We see that the evolution of $\Phi_{\rm spin}$ shows a non-uniform decreasing trend with time.
This can be attributed to the high orbital eccentricity ($e\sim 0.65$) of the massive BH binary in OJ~287, which ensures comparatively stronger spin-orbit interactions during pericentre passages resulting in rapid changes in $\Phi_{\rm spin}$.
This implies that, in our \textit{spin model}, the radio jet of OJ~287 will experience wobbling following the spin precession of the primary BH.

We now move on to explore various implications of the \textit{disc model}.

\begin{figure*}
    \centering
    \includegraphics[width=0.45\textwidth]{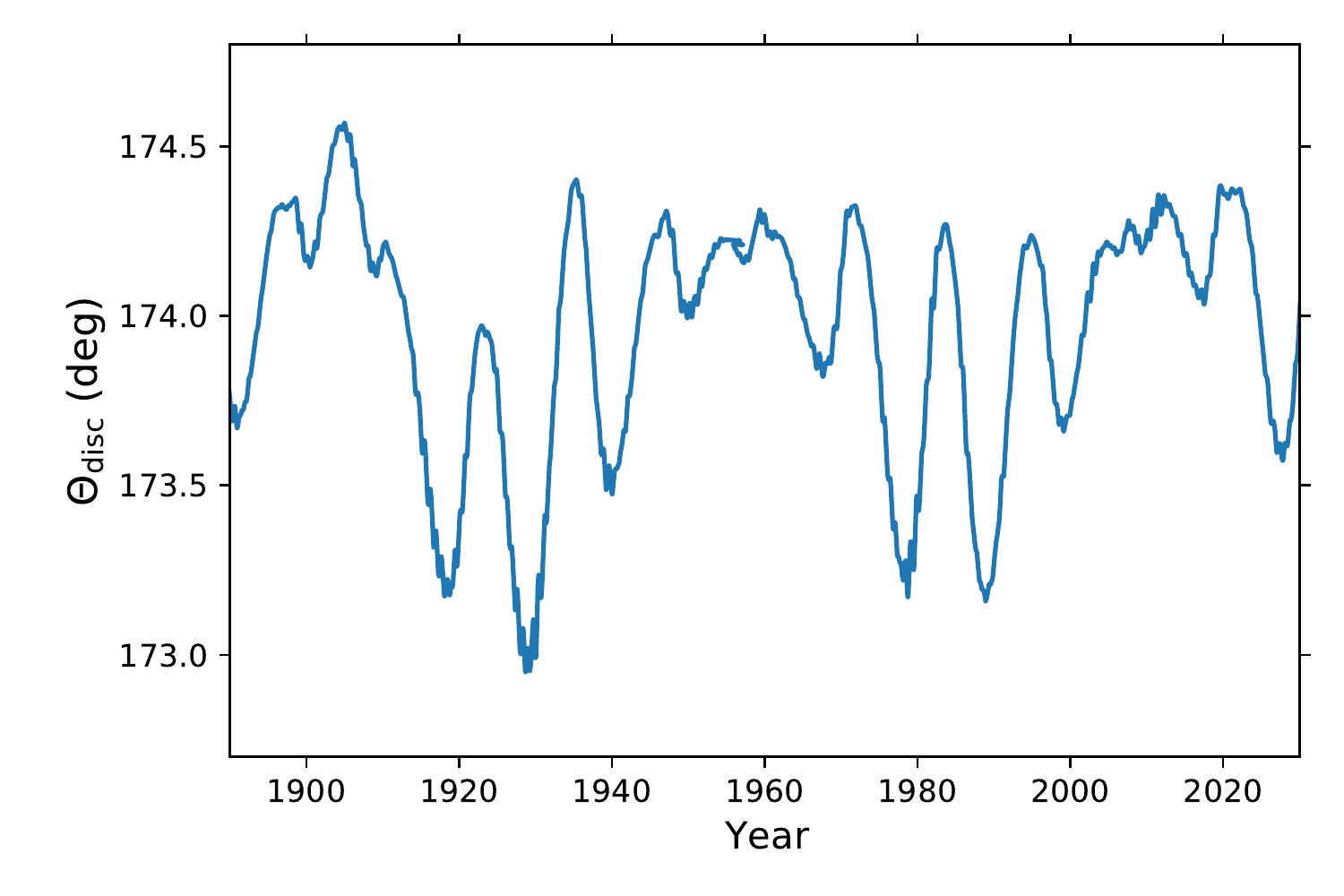}
    \includegraphics[width=0.45\textwidth]{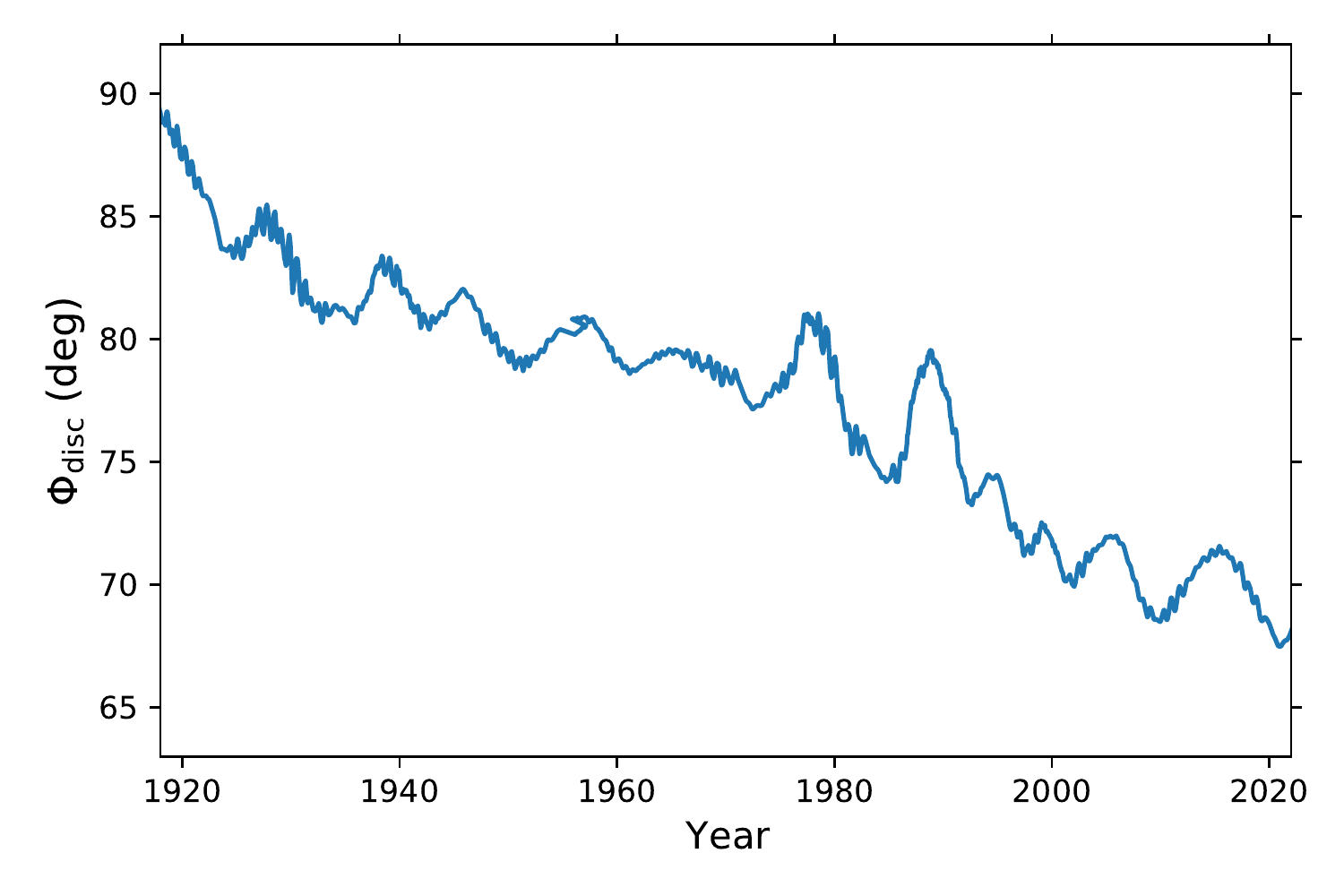}
    \caption{Temporal evolution for the orientation of the average angular momentum of the inner region of the accretion disc.
    The left panel shows the time evolution for the polar angle $\Theta_{\rm disc}$ while the right panel provides its azimuthal counterpart $\Phi_{\rm disc}$. The variations in $\Theta_{\rm disc}$ are caused by the gravitational interaction of the secondary BH with the disc particles while $\Phi_{\rm disc}$ mainly follows the precession of the spin of the primary BH.}
    \label{fig:disc_prec}
\end{figure*}

\subsection{Accretion disc model for OJ~287's jet}
\label{sec:disc_model}

The \textit{disc model} involves tracking the evolution of the angular momentum direction of the inner region of the accretion disc as it determines the radio jet direction.
This requires us to model the simultaneous evolution of the accretion disc in the presence of our BH binary.
To achieve this, we adapt and extend the BBH accretion disc evolution model of \citet{pih13}, where  the accretion disc consists of a cloud of point particles moving under the influence of viscous forces while interacting with the two BHs gravitationally.
The accretion disc is divided into a radially non-uniform polar grid, and  a particle in a grid cell can only interact with  other particles present in the same cell \citep{mil76}.
The viscous forces acting on a particle are given by 
\begin{eqnarray}
    F_i = \nu \, (\mathrm{v}_{i} - \Bar{\mathrm{v}}_i) \,,
    \label{eqn:viscous_force}
\end{eqnarray}
where $\nu$ is the kinematic viscosity at that region while $\mathrm{v}_i$ and $\Bar{\mathrm{v}}_i$ stand for the velocity component of a disc particle in a cell and the mean velocity component of all the particles present in that cell, respectively. 
In practice, we calculate these viscous forces only in the radial and vertical directions \citep{pih13}.
Further, we calculate the viscosity by employing the $\alpha_g$ disc model following \citet{LV96}. This model is a variant of the thin accretion disc model of \citet{SS73}, where the presence of a magnetic field provides stability to the disc \citep{SC81}.
In our model, the gas pressure and magnetic pressure are in equilibrium, and  radiation pressure dominates over both of them in the inner region. 
Let us note that the disc properties are uniquely determined by the central mass, the accretion rate of the primary BH and the viscosity coefficient $\alpha_g$. 
We approximate the central mass to be that of the primary BH ($\sim 18 \times 10^9 M_{\odot}$), while  the accretion rate is extracted from the total un-beamed luminosity of OJ~287 \citep{worrall82}, which turns out to be $\Dot{m} \sim 0.1 \, \Dot{m}_{\rm Edd}$. We have assumed $\alpha_g = 0.1$, which is a reasonable value for an AGN accretion disc  \citep{king07,HK01}.

To model OJ~287's accretion disc, we distribute uniformly $50,000$ particles from $3 R_s$ to $100 R_s$ (the Schwarzschild radius $R_s \sim 362$ AU for the primary BH).
Initially, all particles are in circular orbits around the primary BH and reside on the $X-Y$ plane.
The initial velocity of the disc particles are given as $\mathbf{v} = \mathbf{\omega} \times \mathbf{r}$ where we use PN accurate values of the angular velocity $\mathbf{\omega}$ for circular orbits \citep{bla14}. 
The whole system consisting of the two BHs and all the disc particles is simultaneously evolved.
As noted earlier, we incorporate gravitational interactions between the two BHs, between BHs and disc particles, and viscous forces to characterise particle-particle interactions in the disc.
For gravitational interaction, we use 3PN accurate orbital dynamics with leading order spin-orbit interaction, leading order radiation reaction and classical quadrupolar interaction.
While evolving the BBH-accretion disc system, we follow the three Cartesian components of the position and the velocity vectors of every particle present in the accretion disc.
This allows us to specify the angular momentum  direction of each particle using a pair of polar and azimuthal angles, $\Theta_i$ and $\Phi_i$, at every epoch.
It turns out that the values of these angles generally do not vary significantly within a cell or between neighbouring cells.
Thereafter, we follow the temporal variations in $\Theta_i$ and $\Phi_i$ angles of every disc particle to infer how the disc is evolving over time. 
Interestingly, we observe that the disc particles situated up to $2200$ AU from the central BH show precession of their orbital planes. 
This allows us to obtain the time evolution in the direction of the mean angular momentum of all the particles situated within $2200$ AU from the centre. 

We show in Figure~\ref{fig:disc_prec}  how the orientation of the inner part of our accretion disc changes during a $100$ year period from 1920 to 2020.
The left and right panels of Figure~\ref{fig:disc_prec} display respectively the temporal evolution of the polar  ($\Theta_{\rm disc}$) and  azimuthal  ($\Phi_{\rm disc}$) angles that specify the average angular momentum direction of all disc particles up to $2200$ AU from the primary BH.
The angles $\Theta_{\rm disc}$ and $\Phi_{\rm disc}$ here have the same geometrical interpretation as the angles $\Theta$ and $\Phi$ of Figure~\ref{fig:coordinate}, respectively, except instead of BH spin, $\mathbf{S}_1$ direction now corresponds to the direction of average angular momentum of the inner region of the disc.
In contrast to our spin model, the polar angle $\Theta_{\rm disc}$ does vary in time by a few degrees in the disc model.
This may be attributed to the perturbations experienced by the disc particles due to the nature of the secondary BH trajectory.
Such an explanation is consistent with the presence of two rough periods of $12$ and $60$ years visible in the $\Theta_{\rm disc}$ plot.
These two time scales, inherent in our BBH central engine, are associated with the periods of binary BH orbit and its relativistic advance of the pericentre.
The right panel of Figure~\ref{fig:disc_prec} displays the temporal evolution for the azimuthal angle $\Phi_{\rm disc}$ which shows a systematic decrease in its value during the $100$ year period between 1920 to 2020.
The $\Phi_{\rm disc}$  evolution is not purely secular, but contains an oscillatory component with a periodicity of around 12 years.
We conclude that these variations may be associated with gravitational interactions between the disc particles and the primary BH.
Such  interactions try to align the angular momentum direction of the inner part of the accretion disc with the spin direction of the primary BH.
Recall that in such a \textit{disc model}, the precession of the radio jet is governed by the evolution of polar and azimuthal angles displayed in Figure~\ref{fig:disc_prec}.
In the next section, we make contact with the above results from our two models with the existing observations.

\section{Observational implications}
\label{sec:obs_implication}
In this section, we explore possible observational implications of our BBH central engine model for OJ~287 in the context of high-resolution radio images of the parsec-scale jet.
First, we explore the feasibility of explaining the observed variations in the PA of the radio jet of OJ~287 at the three radio frequencies using our two approaches described in Sections~\ref{sec:bhh_spin} and~\ref{sec:disc_model}.
Additionally, we probe  plausible implications of our explorations during the EHT/GMVA era.

\begin{figure}
    \centering
    \includegraphics[width=0.95\columnwidth]{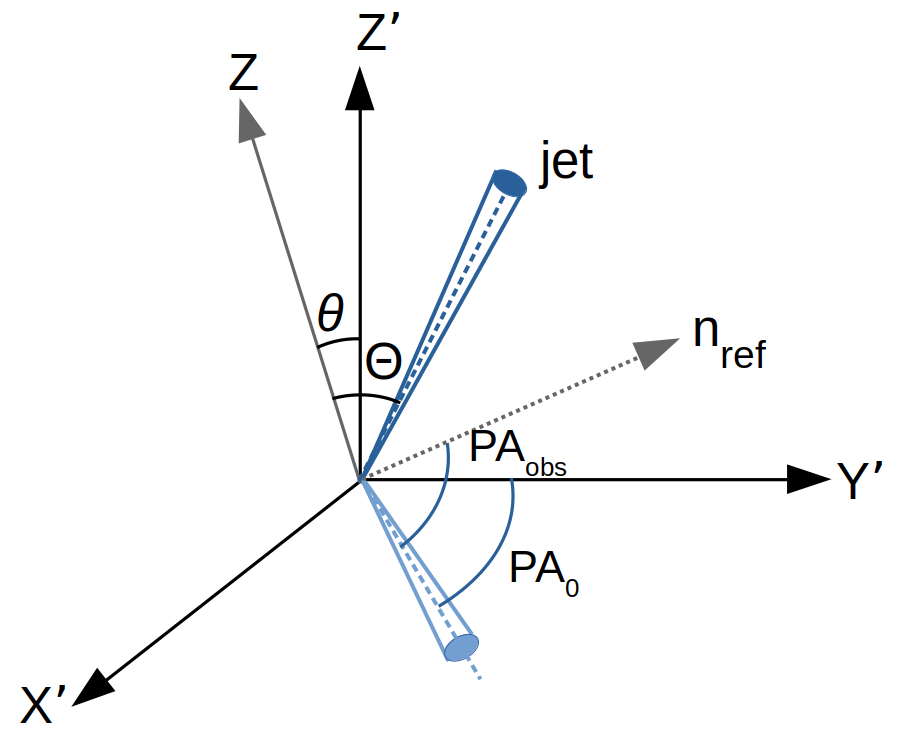}
    \caption{Position angle of the projected jet on the sky plane.
    We let the sky plane to be the $X'Y'$ plane while the $Z'$ axis represents the observer's line of sight and $\rm n_{ref}$ is the reference direction on the sky plane from which the PA of the projected jet is measured. 
    The Z-axis here is the same as the one in the invariant coordinate system $XYZ$ (Figure~\ref{fig:coordinate}) in which the directions of the jet and the observer's line of sight are measured with $[\Theta_{\rm jet}, \Phi_{\rm jet}]$ and $[\theta_{\rm obs}, \phi_{\rm obs}]$, respectively. 
    The angles $\Theta_{\rm jet}$ and $\theta_{\rm obs}$ are respectively denoted by $\Theta$ and $\theta$ in the figure and angles and $\Phi_{\rm jet}$ and $\phi_{\rm obs}$ are not shown to preserve the clarity.
    The angle $\rm PA_{obs}$ indicates the observed PA while $\rm PA_0$ denotes the PA value we calculate in our model.
    A constant offset ($\rm PA_{obs} - PA_0$) may be present between the observed and calculated PA values due to the lack of information about the $\rm n_{ref}$ direction in our coordinate system.
    }
    \label{fig:dpa}
\end{figure}

\subsection{Connecting Jet Position Angle to Jet Direction}
\label{subsec:pa_model}

We begin by denoting the pair of angles that specify the radio jet direction of OJ~287 to be $[\Theta_{\rm jet}, \Phi_{\rm jet}]$.
Obviously, we identify this pair either with $[ \Theta_{\rm spin},\Phi_{\rm spin}]$ or $[ \Theta_{\rm disc},\Phi_{\rm disc}]$ according to the circumstances.
The next natural step then involves connecting the temporal evolution for  $\Theta_{\rm jet}$ and $\Phi_{\rm jet}$, detailed in the previous sections, to the observed variations in the PA of the radio jet of OJ~287.
It turns out that the jet PA depends on the angle between the jet direction and the line of sight of the observer. 
We parametrize the line of sight to the observer direction using two angles $\theta_{\rm obs}$ and $\phi_{\rm obs}$ in our invariant coordinate system (Figure~\ref{fig:coordinate}) to make contact with the existing observations.
Additionally, we need two more parameters to specify the jet PA observed at a particular radio frequency at any given epoch.
The first parameter arises because any changes to the jet direction, presumably due to changes in the central engine, requires a certain time interval to propagate and become observable. 
Such a \textit{time delay} ($\Delta t$) should depend on the frequency at which the jet is observed, as low-frequency  observations sample the jet farther away from its origin as compared to high-frequency radio observations.
This ensures that  $\Delta t$ values at low frequencies are larger than their higher frequency counterparts.

The second parameter ($\Delta PA$) has two components: (i) a geometrical component arising due to the coordinate system and (ii) a physical frequency-dependent component arising from the bending of the jet.
The reason for the presence of the first component is explained in Figure~\ref{fig:dpa}, where the $X'Y'$ plane is the sky plane and the $Z'$ axis represents the observer's line of sight.
The PA of the projected jet is measured from a fixed reference direction ($\rm n_{ref}$ in Figure~\ref{fig:dpa}) on the sky plane.
However, the angles $\theta_{\rm obs}$ and $\phi_{\rm obs}$, which are used to define the observer’s line of sight ($Z'$) w.r.t.\ the invariant frame ($XYZ$ frame in Figure~\ref{fig:coordinate}), only determine the plane of the sky and not the direction ($\rm n_{ref}$) from which the PA is measured. 
Therefore, the PA values that we measure in our coordinate system are expected to have a constant offset ($\rm PA_{obs} - PA_0$) from the actual values which should be independent of the observation frequency.
Additionally, the jet may also bend as it propagates, such as in the helical geometry proposed by \citet{VP13}. This introduces the second component of $\Delta PA$ which should depend on the frequency of observation.

These considerations imply that the PA of the jet at a particular radio frequency at any given epoch $t$ should depend on $\Theta_{\rm jet}$ and $\Phi_{\rm jet}$ calculated at a time $t-\Delta t$, $\Delta PA$ and the direction of the line of sight specified by $\theta_{\rm obs}$ and $\phi_{\rm obs}$. 
To calculate $ PA\,(t)$ we use the following expression 
\begin{widetext}
\begin{equation}
    PA\,(t) = \arctan \left( \frac{\cos{\theta_{\rm obs}} \sin{\Theta_{\rm jet}(t - \Delta t)} \cos{(\Phi_{\rm jet}(t - \Delta t) - \phi_{\rm obs})} - \sin{\theta_{\rm obs}} \cos{\Theta_{\rm jet}(t - \Delta t)}}{\sin{\Theta_{\rm jet}(t - \Delta t)} \sin{(\Phi_{\rm jet}(t - \Delta t) - \phi_{\rm obs})}} \right) + \Delta PA \,,
    \label{eqn:PA_calc}
\end{equation}
\end{widetext}
\noindent 
where $\Theta_{\text{jet}}(t)$ and $\Phi_{\text{jet}}(t)$ provide these angles as functions of time,
with the assumption that the angle between the jet direction and the observer's line of sight is small, applicable to a blazar such as OJ~287.
Clearly, we have two prescriptions for $PA(t)$ due to  \textit{spin} and \textit{disc model} based evolution for $\Theta_{\rm jet}$ and $\Phi_{\rm jet}$.

\subsection{ Modelling of Multi-epoch Multi-frequency Jet PA observations}
\label{subsec:results}

\begin{figure*}
\centering
\begin{tabular}{c}
     \includegraphics[width=0.85\textwidth]{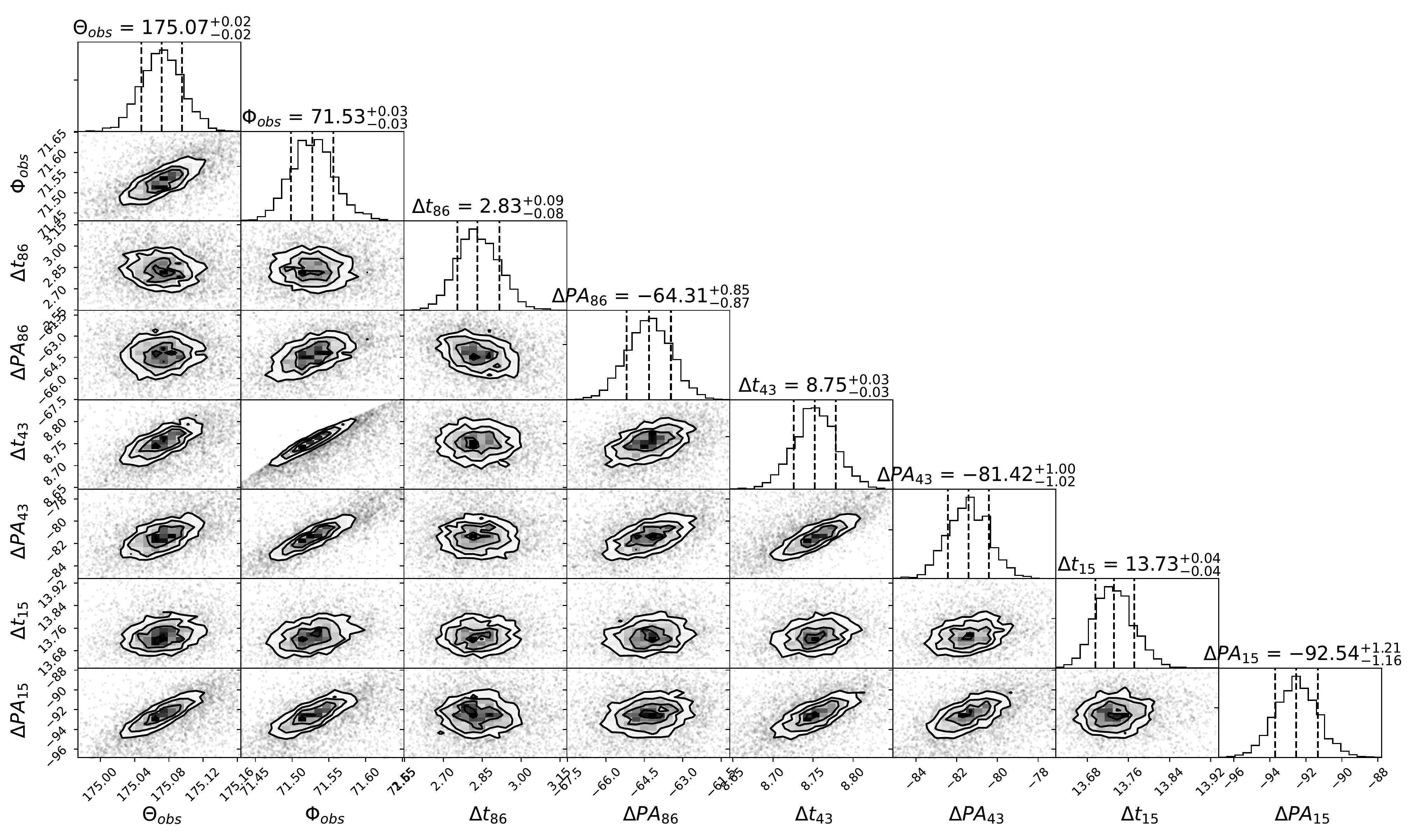}
     \\
    (a) Spin model\\  
    \includegraphics[width=0.85\textwidth]{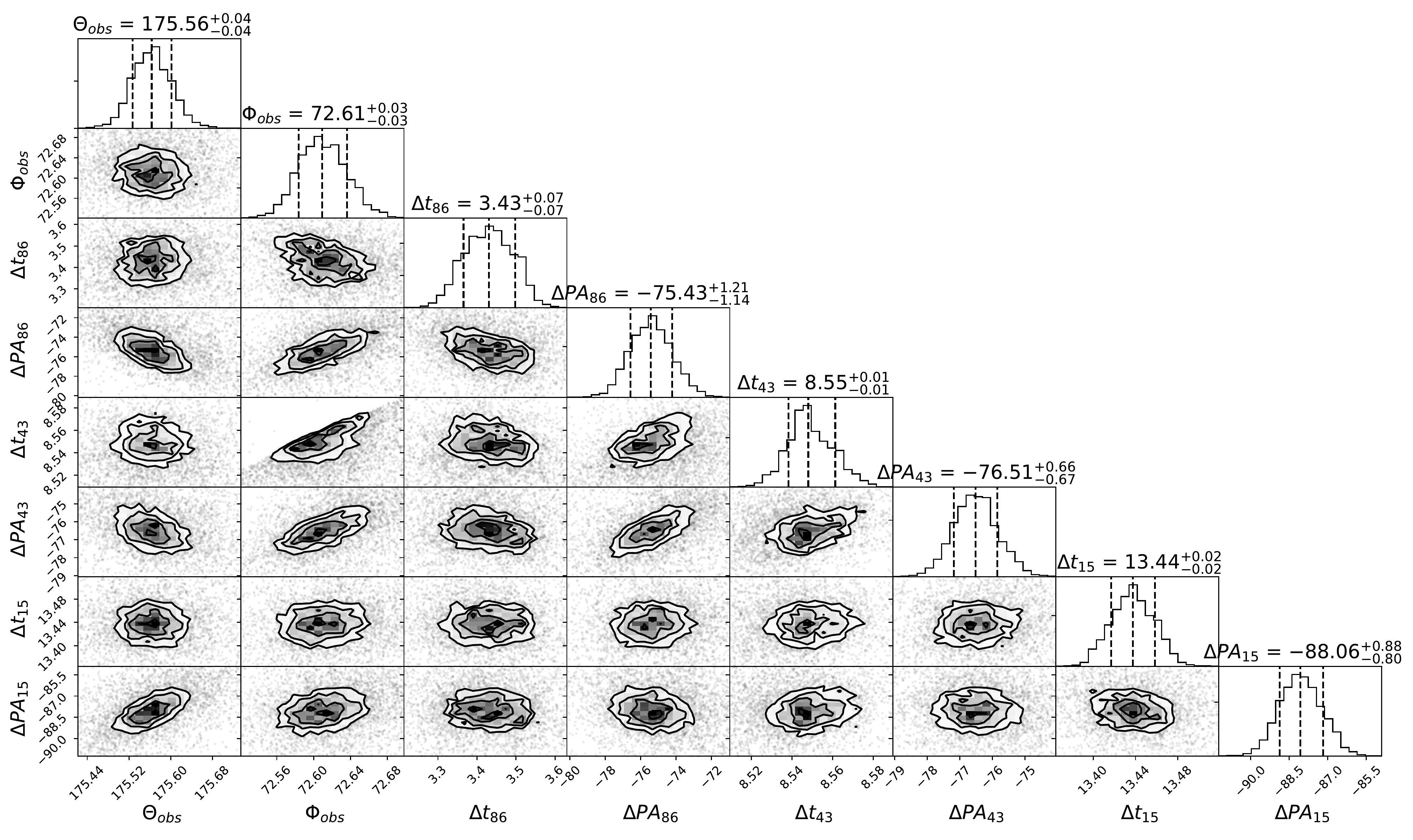}\\
    (b) Disc model
\end{tabular}
    \caption{The marginalised posterior distributions of the parameters $\boldsymbol{\hat\Theta}$ computed using PA data at three radio frequencies while employing the jet precession arising from (a) the \textit{spin model} and (b) the \textit{disc model}.} 
    \label{fig:posterior}
\end{figure*}

\begin{table*}
	\centering
	\caption{The prior distribution for the eight model parameters and median values of the fitted parameters with their 1-$\sigma$ uncertainties for both the \textit{spin model} and \textit{disc model}. $U[a,b]$ represents a uniform distribution between $a$ and $b$.. The uncertainties are calculated from the posterior distribution of our parameters. The natural log values of the evidences for both the models are given in the bottom row.}
	\label{tab:param_posterior}
	\begin{tabular}{lclrr} 
		\hline
		Parameters & Unit & Prior Distribution & Spin Model & Disc Model\\
		\hline
		$\theta_{\rm obs}$ & deg & U[170,180] & $175.07 \pm 0.02$ & $175.56 \pm 0.04$\\
		$\phi_{\rm obs}$  & deg & U[65,80] & $71.53 \pm 0.03$ & $72.61 \pm 0.03$\\
		$\Delta t_{86}$  & year & U[0,5] & $2.83^{0.09}_{-0.08}$ & $3.43 \pm 0.07$\\
		$\Delta PA_{86}$  & deg & U[-100,-50] & $-64.31^{+0.85}_{-0.87}$ & $-75.43^{+1.21}_{-1.14}$\\
		$\Delta t_{43}$  & year & U[5,10] & $8.75 \pm 0.03$ & $8.55 \pm 0.01$\\
		$\Delta PA_{43}$  & deg & U[-100,-50] & $-81.42^{1.00}_{-1.02}$ & $-76.51^{+0.66}_{-0.67}$\\
		$\Delta t_{15}$  & year & U[10,16] & $13.73 \pm 0.04$ & $13.44 \pm 0.02$\\
		$\Delta PA_{15}$  & deg & U[-110,-70] & $-92.54^{+1.21}_{-1.16}$ & $-88.06^{+0.88}_{-0.80}$\\
		\hline
		ln$\mathcal{Z}$ &   &  & -1133 & -1543\\
		\hline
	\end{tabular}
\end{table*}

We now simultaneously fit the jet PA of OJ~287 observed at three different frequencies (86 GHz, 43 GHz and 15 GHz) with our two models. 
From equation (\ref{eqn:PA_calc}), we identify eight independent fitting parameters: two angles ($\theta_{\rm obs}$ and $\phi_{\rm obs}$) which define the observer's line of sight, and three sets of two parameters ($\Delta t$ and $\Delta PA$) for each frequency of observation. 
The parameters $\Delta t$ and $\Delta PA$  for different frequencies are denoted with the frequency in GHz as a subscript: $\Delta t_{86}$ and $\Delta PA_{86}$ for 86 GHz, $\Delta t_{43}$ and $\Delta PA_{43}$ for 43 GHz, and $\Delta t_{15}$ and $\Delta PA_{15}$ for 15 GHz.
We collectively denote this set of eight parameters by ${\boldsymbol{\hat\Theta}}$.
We note that the $\Delta t$s at different frequencies should not be independent and are expected to be some function of the frequency. 
Unfortunately, there is no well defined or accurate model for the dependency of the time delays on the observational frequencies. 
Therefore, we used three different $\Delta t$s for three different frequencies as mentioned above to perform our fitting in a model-independent way. 
We explore the dependency of $\Delta t$ on observation frequency from our results later in this section.

We employ the Bayesian inference technique to estimate the parameters ${\boldsymbol{\hat\Theta}}$ for our two models, namely the \textit{spin model}  and the \textit{disc model}, from the multi-frequency PA measurements.
We provide here a summary of the Bayesian inference that we employ. 
A detailed introduction to Bayesian inference may be found in, e.g., \citet{Hogg2010}.
For the present effort, the Bayes theorem reads 
\begin{equation}
    P[\boldsymbol{\hat\Theta}|D,M] = \frac{P[D|\boldsymbol{\hat\Theta},M] P[\boldsymbol{\hat\Theta}|M]}{P[D|M]}\,,
\end{equation}
where $P[\boldsymbol{\hat\Theta}|D,M]\equiv P_M[\boldsymbol{\hat\Theta}]$ is the posterior distribution of the parameters $\boldsymbol{\hat\Theta}$ assuming the model $M$, $P[D|\boldsymbol{\hat\Theta},M]\equiv \mathcal{L}_M[\boldsymbol{\hat\Theta}]$ is the likelihood function, $P[\boldsymbol{\hat\Theta}|M]\equiv \Pi[\boldsymbol{\hat\Theta}]$ is the prior distribution of the parameters $\boldsymbol{\hat\Theta}$, and $P[\boldsymbol{\hat\Theta}|M]\equiv \mathcal{Z}_M$ is the Bayesian evidence of the model. 
The Bayesian evidence can be considered as a normalization factor which normalizes the product $\mathcal{L}_M[\boldsymbol{\hat\Theta}]\; \Pi[\boldsymbol{\hat\Theta}]$, and can be written as
\begin{equation}
    \mathcal{Z_M} = \int d^N\boldsymbol{\hat\Theta}\; \mathcal{L}_M[\boldsymbol{\hat\Theta}]\; \Pi[\boldsymbol{\hat\Theta}]\,,
\end{equation}
where $N$ is the number of parameters  ($N=8$ in our case).
The marginalised posterior distributions for each parameter $\Theta_i$ in $\boldsymbol{\hat\Theta}$ can be computed by integrating the posterior distribution $P_M[\boldsymbol{\hat\Theta}]$ over the $N-1$ parameters other than $\Theta_i$ (denoted by $\boldsymbol{\hat\Theta}'_i$):
\begin{equation}
    P_M[\Theta_i] = \int d^{N-1} \boldsymbol{\hat\Theta}'_i \; P_M[\boldsymbol{\hat\Theta}]\,.
\end{equation}
Traditionally, one compares two models against available observations by comparing the associated evidence, and the model that gives higher Bayesian evidence is interpreted to be more favoured by the data.
However, we will practice caution while comparing our models with existing multi-frequency observations due to a few additional considerations that will be explained later.

We now proceed to define the likelihood function relevant for fitting our multi-frequency PA measurements. 
Our data consist of PA measurements $PA_{\nu i}$, measured at three frequencies $\nu$ at times $t_{\nu i}$ with uncertainties $\sigma_{\nu i}$.
Assuming the uncertainties $\sigma_{\nu i}$ to be normally distributed, we can write the likelihood function as 
\begin{equation}
    \mathcal{L}_{M}[\boldsymbol{\hat{\Theta}}]\propto\exp\left[-\frac{1}{2}\sum_{\nu}\sum_{i}\left(\frac{PA_{\nu i}-PA_{\nu M}\left(t_{\nu i};\boldsymbol{\hat{\Theta}}\right)}{\sigma_{\nu i}}\right)^{2}\right]\,,
\end{equation}
where $PA_{\nu M}\left(t_{\nu i};\boldsymbol{\hat{\Theta}}\right)$ represents the $PA(t)$ function defined in equation (\ref{eqn:PA_calc}) specialised for the observation frequency $\nu$ while using the model $M$, and  $\sum_\nu$ represents summation over the three observation frequencies.

\begin{figure*}
    \centering
    \includegraphics[width=0.45\textwidth]{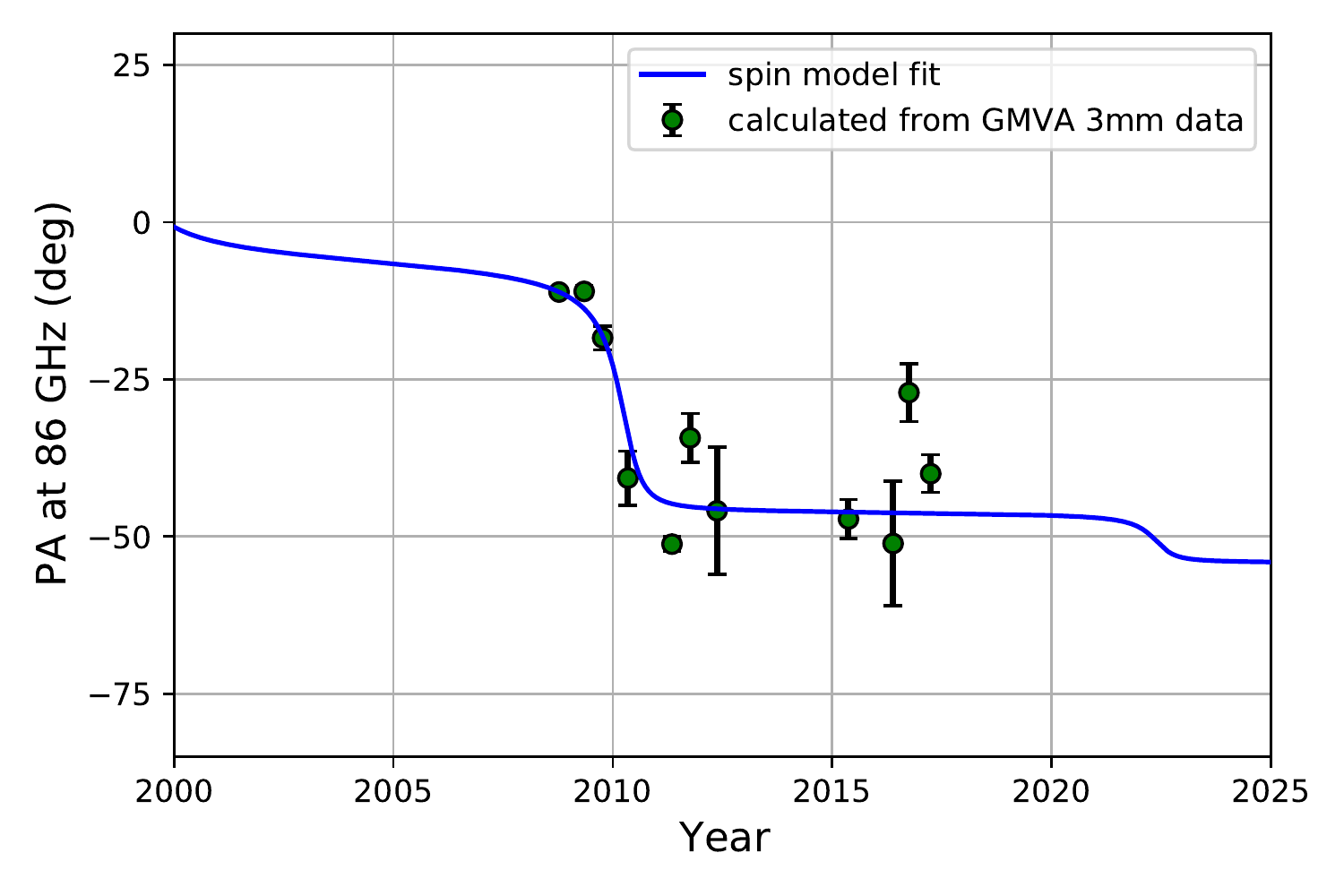}
    \includegraphics[width=0.45\textwidth]{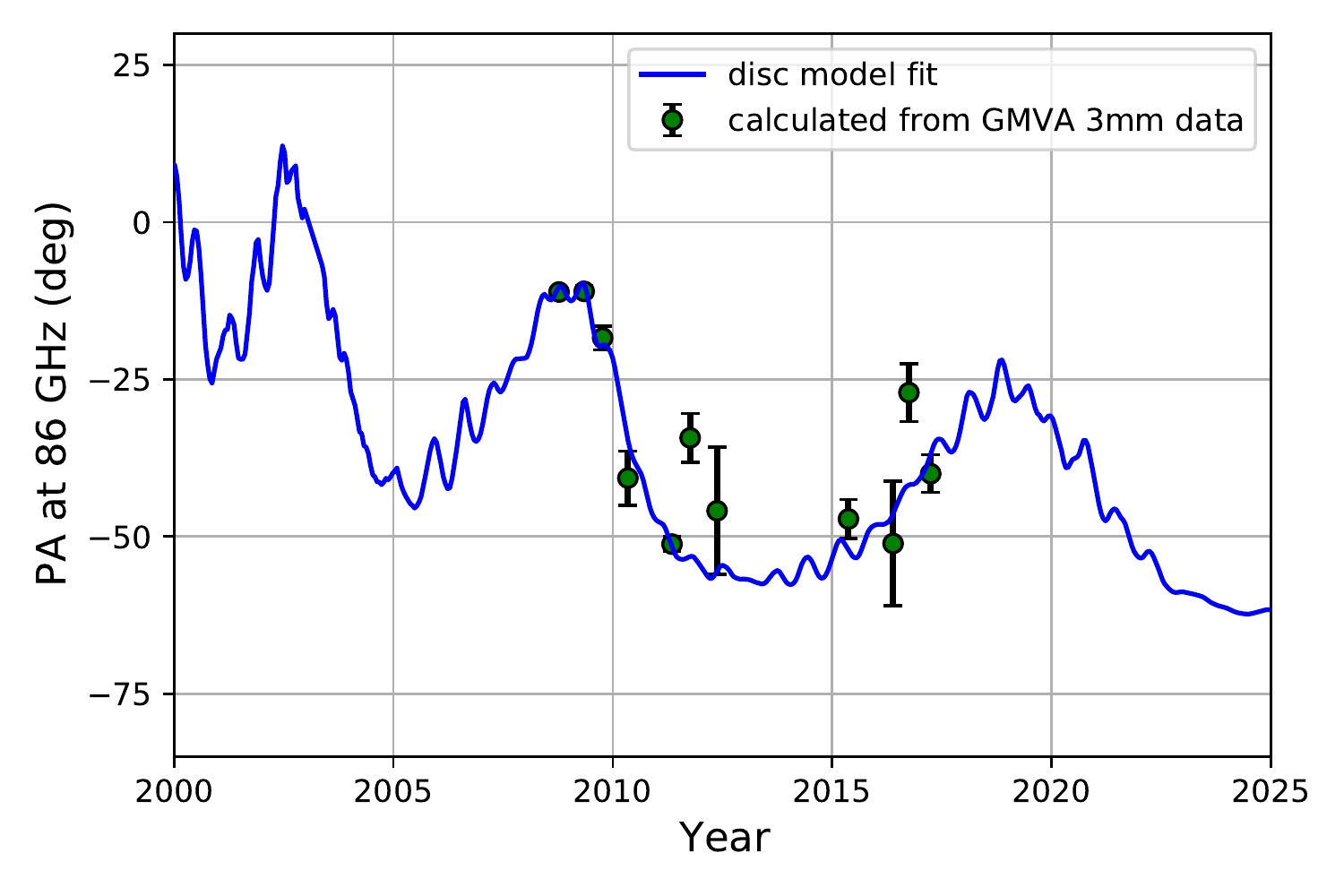} \\
    \includegraphics[width=0.45\textwidth]{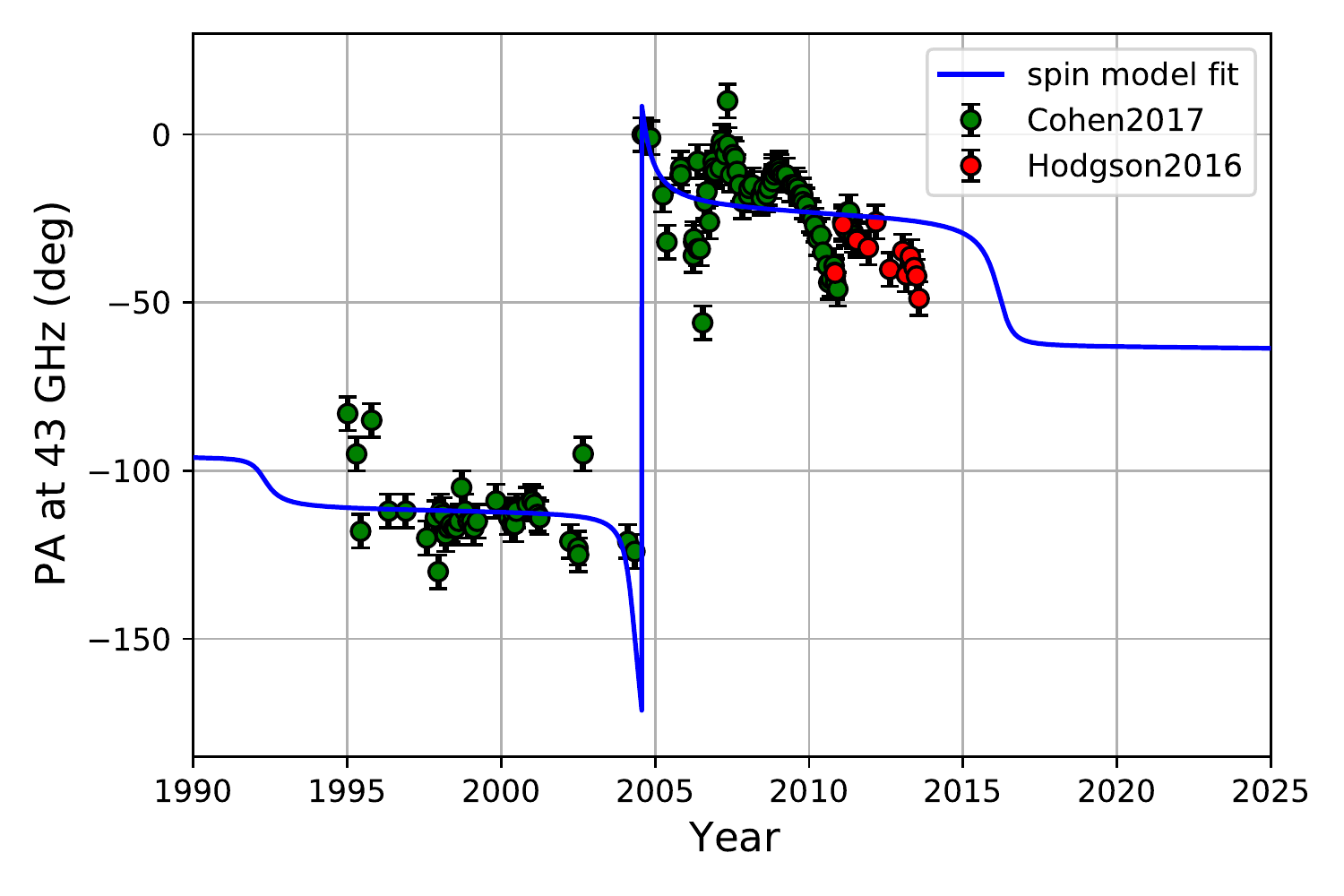}
    \includegraphics[width=0.45\textwidth]{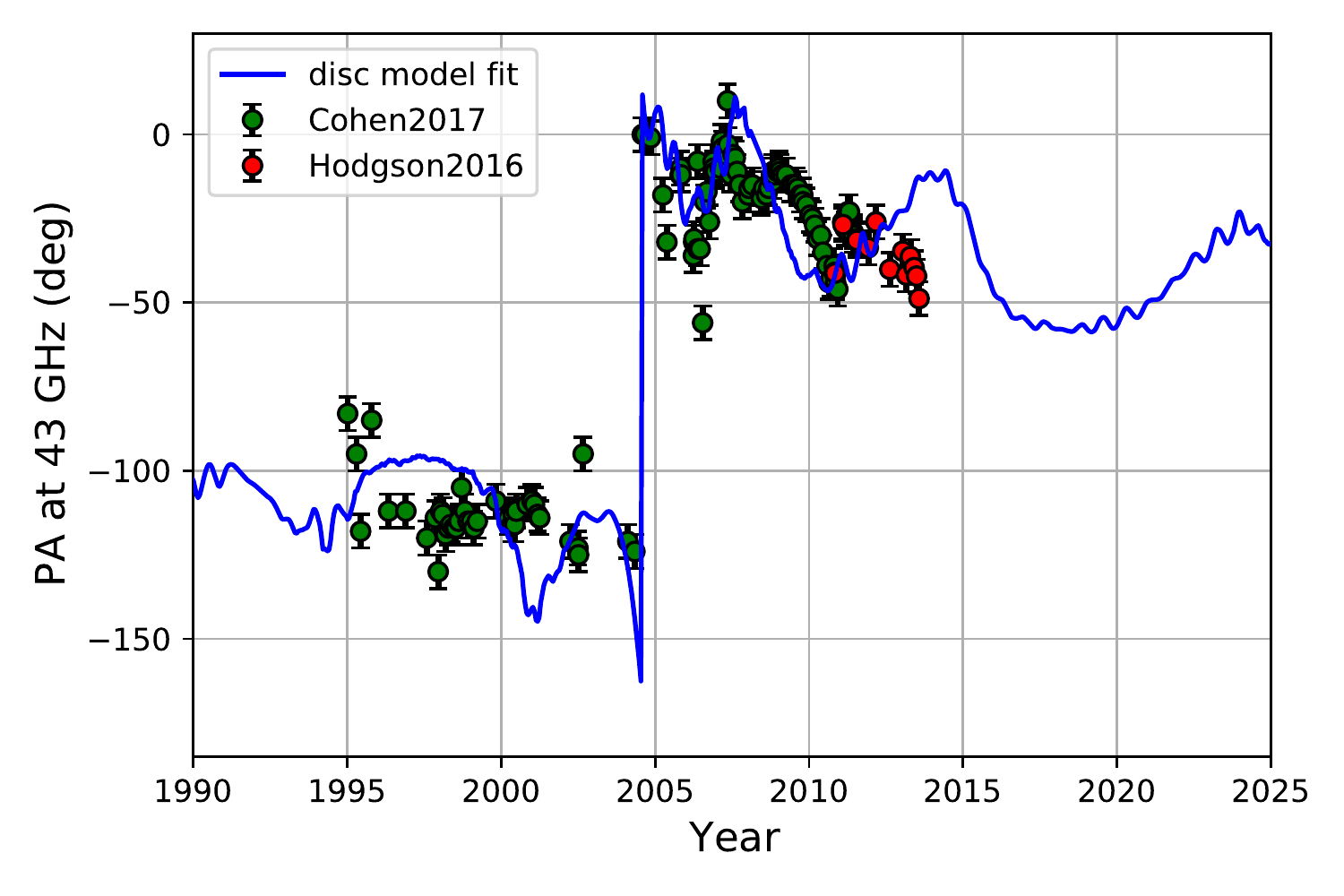} \\
    \includegraphics[width=0.45\textwidth]{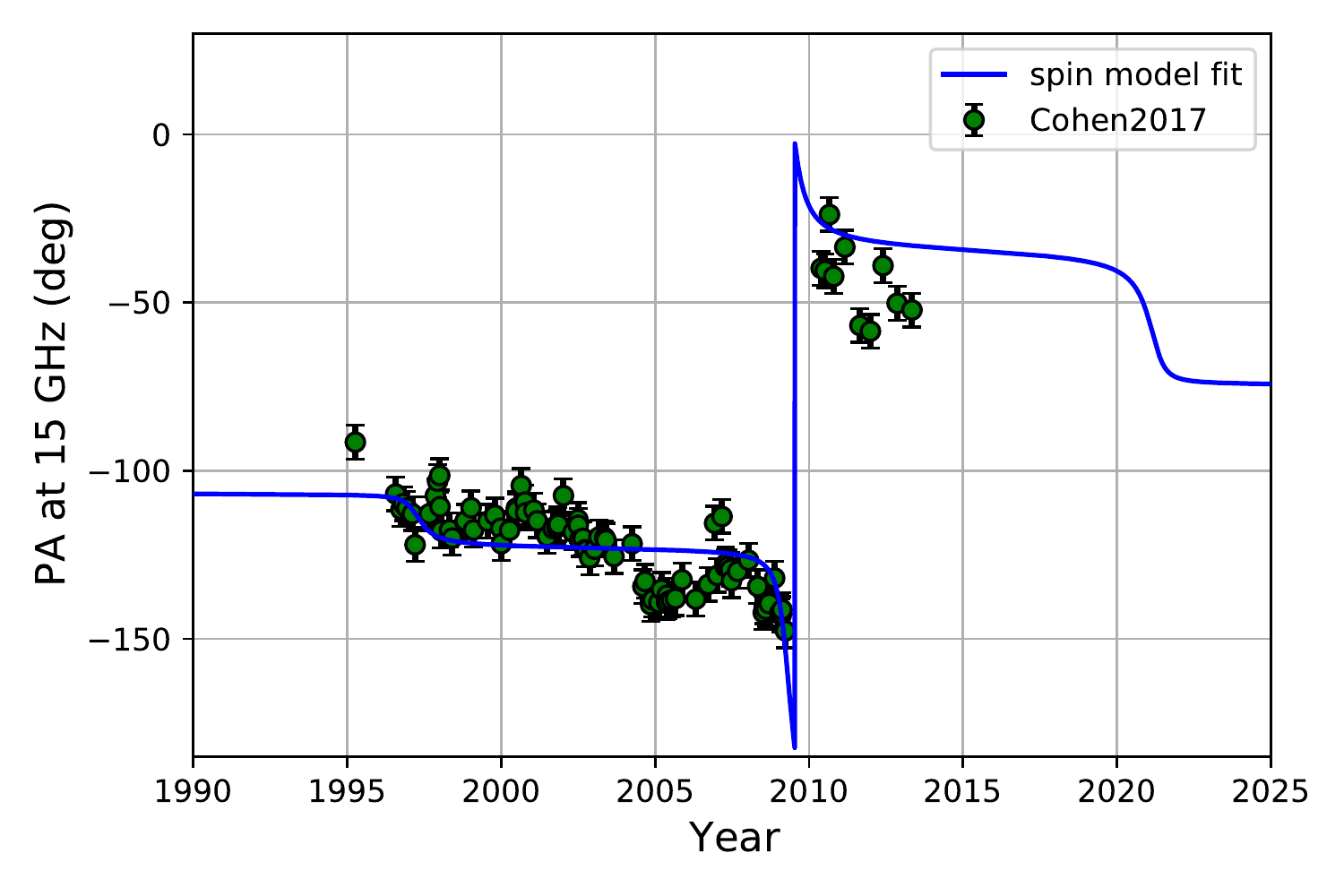}
    \includegraphics[width=0.45\textwidth]{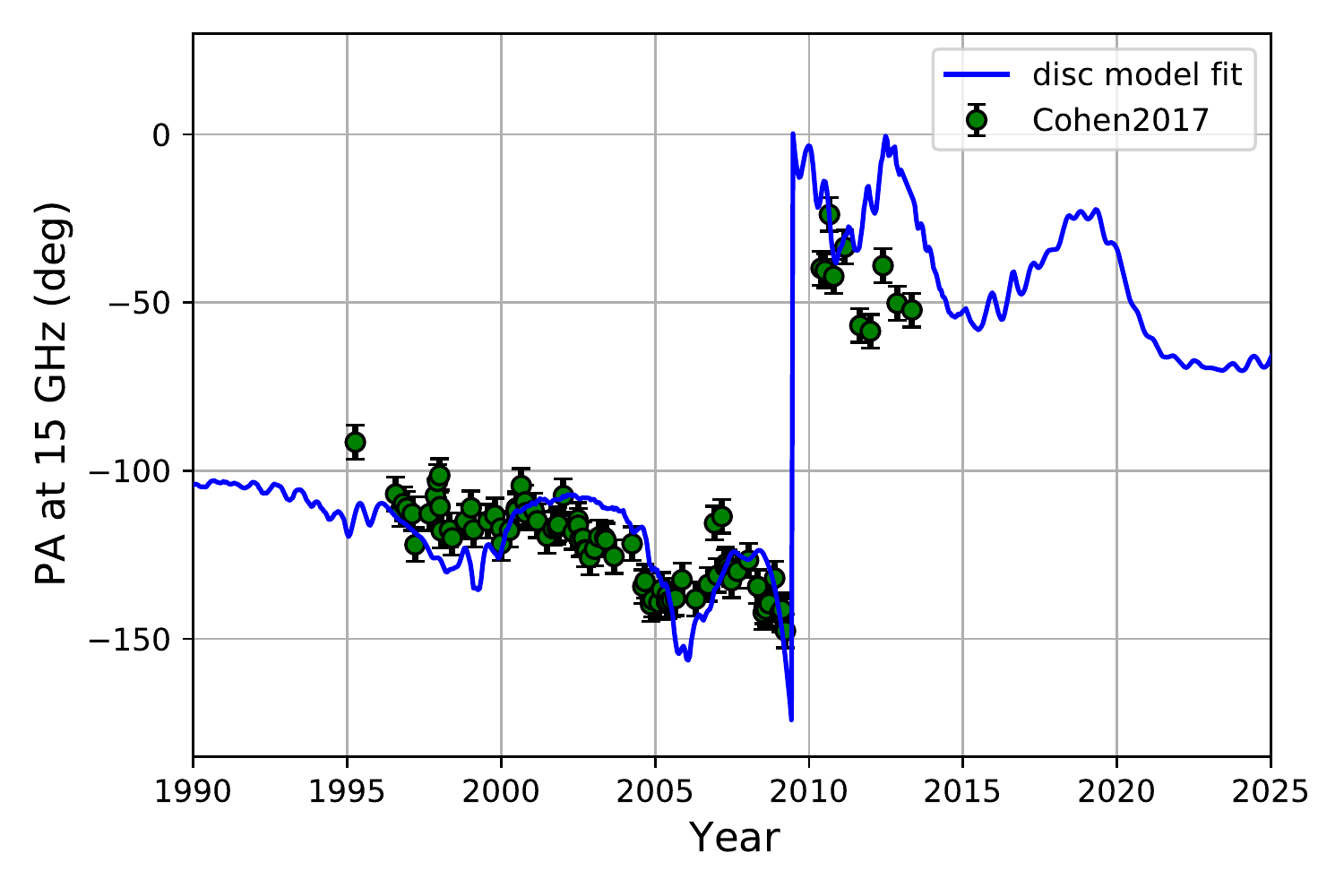} 
    
    \caption{Fits to the variations in the PA of the radio jet at different frequencies with our two jet precession models. The three left panels show the fits with jet precession from the \textit{spin model} and the right panels show the fits with jet precession according to \textit{disc model}. The fit for PA data at 86 GHz is shown in the top panels while the middle ones are for 43 GHz and the bottom ones for 15 GHz. In all plots, green circles with error-bar represent the actual observed PA data and the blue lines are the fit to the data with the models.
    }
    \label{fig:pa_variation_fit}
\end{figure*}

To estimate the parameters $\boldsymbol{\hat{\Theta}}$ for our two models, we draw samples from the posterior distribution.
To this end, we employ the \texttt{nestle}\footnote{Available at \url{https://github.com/kbarbary/nestle}.} package which implements the Nested Sampling algorithm \citep{Skilling2004,Feroz2009}.
This package provides the posterior samples and computes the Bayesian evidences $\mathcal{Z}_M$ relevant for model comparisons.
Further, we use broad uniform priors as listed in Table~\ref{tab:param_posterior}.
We have verified that these priors are wide enough and cause no noticeable impacts on our posteriors.
The marginalised posterior distributions for the \textit{spin model} and the \textit{disc model}, computed from the posterior samples, are displayed  in Figure~\ref{fig:posterior}.
The median and $1\sigma$ credible interval for each parameter are listed in Table~\ref{tab:param_posterior} for both the \textit{spin} and \textit{disc models} along with their Bayesian evidences while employing radio data at three frequencies.

We gather from these results that the direction of observer's line of sight essentially remains the same for both the \textit{spin} and \textit{disc models} (the  average values for $\theta_{\rm {obs}} \sim 175.3^{\circ}$ and $\phi_{\rm {obs}} \sim 72^{\circ}$).
However, the mean \textit{time delay} ($\Delta t$) for different frequencies are different, $\Delta t_{86}$ being the lowest and $\Delta t_{15}$ being the highest.
This is consistent with the fact that the higher frequency observations probe a region closer to the origin of the radio jet compared  with lower frequency observations.
To model the dependency of $\Delta t$ on the observation frequency, we have fitted the obtained $\Delta t$s with a power law of frequency ($\nu$): $\Delta t \propto \nu^{\alpha}$. After fitting with $\Delta t$s for 86, 43 and 15 GHz, we get $\alpha = - 0.66 \pm 0.26$ for the \textit{spin} model and $\alpha = -0.61 \pm 0.20$ for the \textit{disc model}.
Interestingly, the extracted values of $\Delta PA$ are also different for different frequencies and it may be attributed to the bending of the radio jet as discussed above.

Let us now display in Figure~\ref{fig:pa_variation_fit} the fits to the observed PA variations in our three frequency dataset while using  the median values estimated from the marginalised posterior distributions and listed in Table~\ref{tab:param_posterior}.
The three left panels show fits to the \textit{spin model} with the fit for 86 GHz PA data at the top, 43 GHz in the middle and 15 GHz at the bottom. 
Similarly, the right panels are for the \textit{disc model} with the same ordering in frequency.
The Bayesian fits to the \textit{spin model} are smooth, whereas the \textit{disc model} shows short timescale variations. 
The long-term behaviour of PA variations in both models is broadly consistent with the available observations including the sudden jumps in the PA values. 
While comparing the Bayesian evidence values for these two models, listed in Table~\ref{tab:param_posterior}, we see that the \textit{spin model} gives larger evidence implying that the combined datasets prefer the \textit{spin model} over the \textit{disc model}.
However, it should be noted that the Bayesian evidence values are dominated by the more numerous lower frequency (15 and 43 GHz) observations, for which reliable error bars are not available. 
We use uniform 5$^\circ$ error bars for these data points following \citet{hodgson17}, as mentioned in Section~\ref{sec:pa_data}. 
It is possible that we are underestimating these error bars, which may be causing the large difference in the Bayesian evidence values for our two models. 
Also, it is plausible that the low-frequency observations do not accurately track the direction of OJ 287’s jet or its evolution, which are possibly influenced by other astrophysical effects that may be present as the jet propagates away from the central engine. 
In contrast, for the high-frequency (86 GHz) observations, which probe the jet closer to its origin, we have fewer number of data points.
Therefore, given the data available at the present epoch and above astrophysical considerations, we feel that there is no clear and robust evidence to favour one model over the other.

Interestingly, the predicted PA variations of OJ~287 at different frequencies show different trends for the \textit{spin} and \textit{disc models} in the coming years. 
According to the disc model, PAs at different frequencies should show persistent variations with 12-year timescales. 
But, in the spin model, we do not expect such trends and the PA variations are predicted to be quite flat. 
Therefore, the ongoing and upcoming high-frequency and high-resolution VLBI observations may allow us to identify the more favourable model to explain OJ~287's observed PA variations.
In what follows, we probe the possible implications of our efforts in the context of EHT campaigns on OJ~287.

\subsection{Implications for the on-going EHT campaigns on OJ~287}
\label{subsec:EHT}

OJ~287 has been observed by the EHT in the 2017 and 2018 campaigns, in combination with quasi-simultaneous GMVA+ALMA and space-VLBI RadioAstron observations, with the aim to test our BBH model for OJ~287. Further GMVA+ALMA observations in 2019, 2020, and planned observations for the coming years aim to determine the innermost PA of the primary jet for comparison with the BBH model predictions.
These considerations prompted us to estimate the primary BH jet orientations for the past and future EHT observational epochs at 230 GHz and to probe implications if the  secondary BH develops a jet due to its impacts with the primary BH accretion disc.

We specify the primary BH jet orientations by estimating the expected radio jet PA values at 230 GHz at various epochs. 
We gather from Section~\ref{subsec:results} that the time delay ($\Delta t$) decreases with increasing observational frequency and the $\Delta PA$ values also depend on the frequency.
Therefore, it is reasonable to expect that the PA values at 230 GHz follow the existing and predicted trends at $86$GHz, shifted backwards in time by $2-3$ years with an unknown vertical shift.
Further, it should be possible to provide observational constraints on our $\Delta t$ and $\Delta PA$ at the 230 GHz frequency using PA values, extracted from the 2017 and 2018 EHT campaigns on OJ~287.
With these observational constraints, we can provide  predictions for the orientation of OJ~287's radio jet on the sky plane for the near-future EHT observations.
We may predict the expected PA values at 230 GHz, similar to the way we predicted the possible PA values that GMVA campaigns can measure at 86 GHz in the near future.

Additionally, the EHT campaigns are capable of observing/resolving a secondary jet if the secondary BH supports a temporary active jet. 
Interestingly, the occurrences of the so-called precursor flares in the optical light curve were associated in the model with the turning on of the secondary jet in OJ~287 \citep{pih13}.
The optical data reveal that such precursor flares occurred during 1993, 2005 and 2012.
Therefore, it is plausible that the secondary BH may sustain a temporary jet, although at present there is no firm observational evidence for the presence of the secondary jet.
But if the secondary jet is activated during the precursor flares, it will appear as a secondary core along with a jet which will first be visible in high-frequency radio observations pursued by the EHT and GMVA consortia. 
Though the secondary core may not be resolvable from the primary core, the secondary jet may be observed as a new jet component and we explore below its possible observational implications.

The exact timescale for the emergence of the secondary's jet depends on a number of unknown factors, including properties and geometry of the accretion disc and its corona, the strength and geometry of the magnetic field, and the density of the interstellar medium (ISM) into which the jet is expanding \citep{giannios2011, tchekhovskoy2014, marscher2018}.
However, among extra-galactic systems, the few observations of the time gap between accretion events and subsequent jet ejection events all point to a timescale between days to months. 
Timescales of days have been observed in stellar tidal disruption events \citep[TDEs,][]{komossa2016}, some of which trigger temporary jets following the stellar accretion \citep[e.g.,][]{burrows2011, zauderer2011}. 
A timescale of several months has been observed in X-rays for the 2020 after-flare of the primary SMBH of OJ~287 \citep{komossa2020}, while \citet{Marscher02} found that the time gap between accretion and jet ejection at 43 GHz radio frequency for the galaxy 3C120 is around 0.1 year.
We therefore expect the delay to be around a month or so if the secondary jet in OJ~287 becomes visible at high frequencies of 230 GHz and 86 GHz.

\begin{figure*}
    \centering
    \includegraphics[width = 0.9 \textwidth]{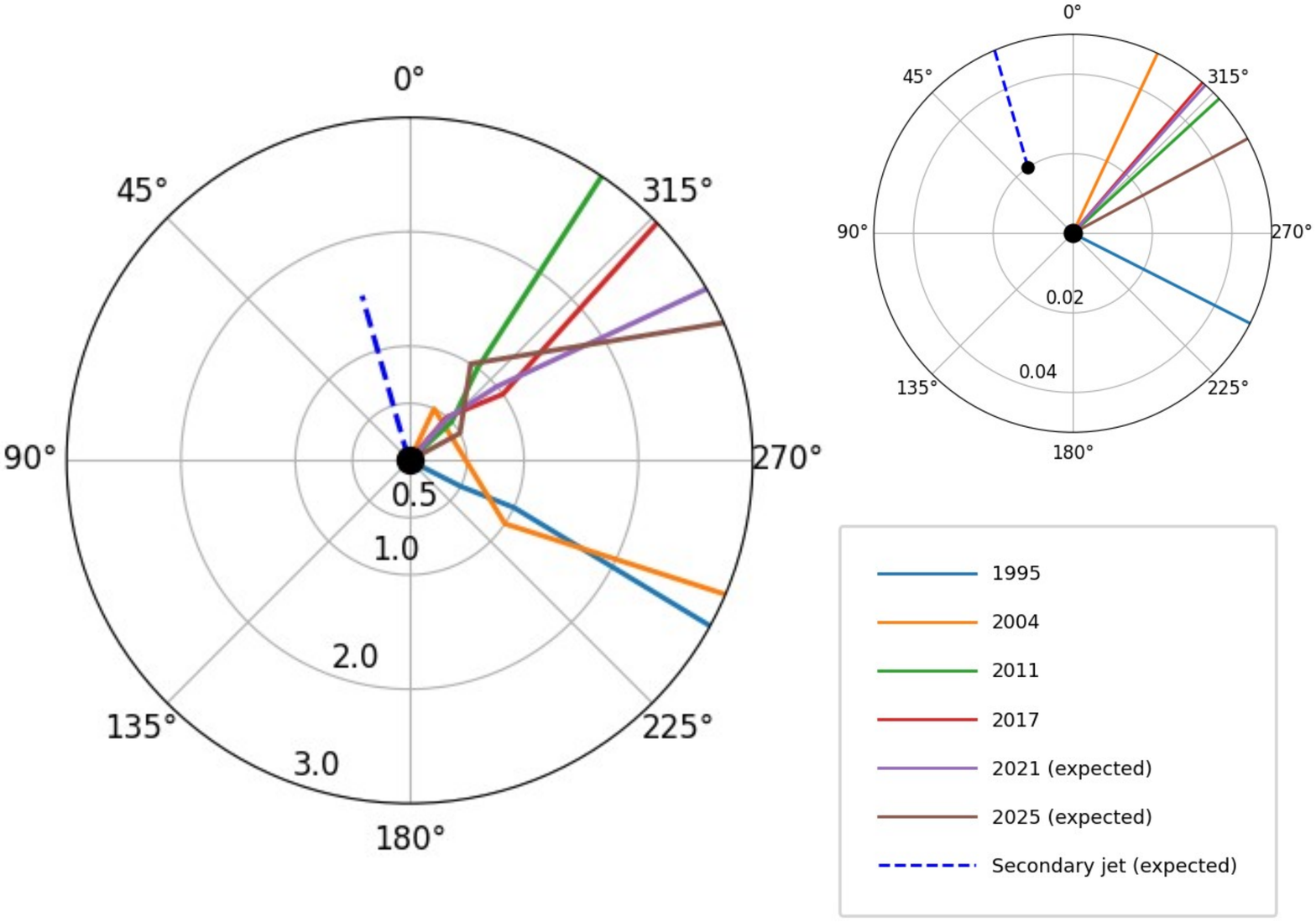}
    \caption{Observed and expected jet directions from the primary and secondary BHs, projected on the sky plane, at different epochs. All the lines, except the dashed blue line, represent the projected primary jet at different epochs. The blue dashed line shows the expected projected secondary jet during early 2021 as explained in the text. The $0^{\circ}$ direction points towards the north and the concentric circles represent 0.5, 1, 2  and 3 mas separation on the sky plane, respectively. At the top right corner, we show a zoomed-in version of the central portion up to 0.05 mas separation from the primary BH. The central (bigger) black dot denotes the position of the primary BH and the smaller black dot indicates the position of the secondary BH on the sky plane during 2013 impact as prescribed in our BBH central engine model.}
    \label{fig:secondary_jet_pa}
\end{figure*}

We now focus on predicting the expected position of the secondary jet, projected on to the plane of the sky, which may appear in a radio image of OJ~287 soon after the precursor and impact flare epochs. 
The parameters from the \textit{disc model} are used to explore the secondary jet orientation on the sky plane.
It is reasonable to expect that the plausible secondary jet will be perpendicular to the disc plane at various impact sites.
The above direction turned out to be essentially a constant in our PN-accurate evolution and not influenced by the orbital phase of the secondary.
This allows us to calculate the PA of the  secondary jet which may appear after the secondary BH impact epochs \citep{pih13}.
In Figure~\ref{fig:secondary_jet_pa}, we display apparent primary and secondary jet directions on the sky plane at various epochs.
The coloured solid lines denote the observed and expected projected primary jet directions on the sky plane at different epochs (we denote the expected primary jet directions in the near future according to the \textit{disc model}). 
In the big circle whose angular radius is roughly 3 mas, the plotted lines are broken as we do include inferences due to all the three radio frequency (86, 43 and 15 GHz) observations. 
The secondary jet direction, expected to appear in high-resolution radio observation of OJ~287,
is marked by the blue dashed line. 
This direction is deduced with the help of parameters present in Table~\ref{tab:param_posterior} and associated with the \textit{disc model}.
To visualise the difficulty in distinguishing the presence of a plausible secondary jet in the central engine of OJ~287, we provide its zoomed-in version in the top right corner of Figure~\ref{fig:secondary_jet_pa} that spans the $0.05$ mas part of its central region. Further, we mark  the position of the 2013 impact point in the zoomed-in region. 
We infer that it will be very difficult to resolve the angular separation between the 2013 impact point and the primary BH position even during the EHT era.
It will be even more difficult to resolve the 2019 and 2022 impact sites as the expected separation is estimated to be much smaller than the 2013 value from the primary BH.

However, it should be possible to resolve the presence of two jets in OJ~287 if they do not point in the same direction. 
The above discussion suggests that these two jets should not point in the same direction. In fact, the secondary jet is expected to point essentially due north in our description. Further, we see that the primary jet is rotated typically by $\sim 90^{\circ}$ from the above direction. 
It is also  worthwhile to consider the analogy to TDEs here, as these also represent temporary high-Eddington-ratio accretion events some of which trigger temporary jets \citep[review:][]{komossa2016}. 
While the inner jet ceases to be powered as the accreting matter diminishes, the outer jet will continue expanding into the surrounding ISM \citep{zauderer2013}. 
TDE jets typically reach their highest radio brightness around a year after the initial disruption/accretion event, consistent with predictions \citep{giannios2011,yang2016}.
If the secondary's radio jet of OJ~287 undergoes a similar evolution, the outer jet will become spatially better resolvable 
since a radial distance of one light year corresponds to a spatial scale of $\sim 67\,\mu$as in OJ~287. 

Finally, it is interesting to point out that the past episodes of the secondary's disc impact would have also triggered temporary radio jet emission.
Evidence for these past transient jet ejections could be searched for in archival deep radio observations of OJ~287.
While a unique association of such features with the secondary might turn out to be difficult, these would appear as individual radio features which deviate in direction and/or kinematics from the main jet of OJ~287.
Typical for these `remnants' from  the past  secondary jet activity 
is that they appear as faint streaks of emission, pointing in a direction different than that of the primary jet, in deep radio images of OJ~287.
These statements may prove helpful if the high-resolution radio images of OJ~287 support the plausible presence of the secondary jet.
The high-frequency radio observations of OJ~287 
by the EHT consortium  may provide our best opportunity to resolve the presence of the primary and secondary jets in this blazar.

\section{Summary and Discussions}
\label{sec:discussion}

We explored the ability of our binary black hole central engine model for OJ~287, developed from its long-term optical observations, to explain the high-frequency radio observations of this unique blazar.
We use the BBH model to describe  OJ~287's observed temporal variations in the PA of its radio jet at 86, 43 and 15 GHz radio frequencies.
We provided two different prescriptions as we do not know for certain what really determines the radio jet directions in active galactic nuclei.
In the \textit{spin model}, we let the jet direction be determined by the primary BH spin in OJ~287 while the \textit{disc model} employed the direction of the angular momentum of the inner region of the accretion disc as its radio jet direction.
Additionally, we employed the existing binary BH central engine parameters, extracted  from its optical observations, in both models while tracking the precession of the jet direction.
A detailed Bayesian analysis reveals that both these models, used for describing the precession of OJ~287's  jet, can broadly explain the observed radio jet PA variations in different radio frequencies.
Moreover, we have provided estimates for the expected PA values for OJ~287 in the coming years, especially at 86 GHz.
These high-frequency observations are expected to follow the actual pointing of the radio jet close to the central engine.
Therefore, it may be possible for us to pinpoint the most favourable description for the radio jet precession in OJ~287 in the near future.
Further, we probed plausible implications of a transitory jet that may emanate from the secondary BH in our BBH central engine model for OJ~287 and provided rough estimates for the primary BH jet direction in the EHT era images of the blazar.

It may be worth listing what the near future observational campaigns might potentially achieve. 
At present, both models are consistent with the long-term temporal variations in the jet PA at different frequencies including the jumps.
There is no clear and robust evidence that the PA data, available at the present epoch, favours one model over the other.
However, our two models predict different trends in PA variations for 86 GHz and 43 GHz observations in the coming years. 
For example, our \textit{disc model} predicts persistent PA variations that increase first and then decrease with a 12-year timescale during the coming epochs, whereas in \textit{spin model} the PA smoothly decreases and then follows a flat line.
The on-going/upcoming GMVA campaigns should be able to distinguish such differences and help us to determine the more favourable model.
Further, it could be worthwhile to contrast the above predictions with what one might expect from a less sophisticated disc model of \citet{VP13}.
In this rather simplistic model, the radio jet is expected to swing further to positive PAs after 2014, to about +20 degrees by 2017 at the 43 GHz resolution (Figure~6 of \citet{VP13}), which is different from what our models predict. 
Therefore, it may even be possible to distinguish  between the two versions of our disc models with the help of continued high-frequency radio observations of OJ~287.
It should be noted that the present model is an improvement over the \citet{VP13} model because we take into account the self-interaction of the disc, and with a larger number of disk particles, are able to concentrate on the region within 6 Schwarzschild radii of the primary where the disc angle variations appear most strongly. We would therefore expect that the present model will follow observations better than the old model.
Therefore, a persistent multi-frequency monitoring of the radio jet from OJ~287 should allow us to further strengthen the presence of a BBH in OJ~287 and to conclude which is the favoured model for its jet direction.

It will also be very interesting, as noted earlier, to substantiate the presence of a secondary jet component in the EHT era radio map of OJ~287.
In our BBH model, the secondary BH can launch a temporary jet after accreting matter during its precursor and impact flare epochs.
The next impact flare is predicted to happen during the middle of 2022 \citep{dey18} and if the secondary jet is activated, the presence of a secondary jet component may emerge in the high-frequency radio map of OJ~287 with a time delay of $\sim$ a month.
Therefore, an appropriate EHT campaign during and after August 2022 should have the best opportunity to observe the possible appearance of the secondary jet in OJ~287.
A similar appearance of the secondary jet is again anticipated during early 2032.
Our present model indicates that the primary and secondary jets, projected on the sky plane, should  point in different directions.
This opens up the possibility of  distinguishing the presence of a secondary jet component in the EHT era radio image of OJ~287.

\section*{Acknowledgements}

We would like to thank the referee for her/his helpful suggestions and detailed comments.
We also thank Marshall H. Cohen for providing us with PA datasets of OJ~287 at different frequencies.
This research has made use of data from the MOJAVE database that is maintained by the MOJAVE team \citep{mojave2018}.
LD thanks FINCA and acknowledges the hospitality of University of Turku. 
LD, AG, and AS acknowledge the  support of the Department of Atomic Energy, Government of India, under Project Identification \# RTI 4002.
JLG and RL acknowledges the support of the Spanish Ministerio de Econom\'{\i}a y Competitividad (grants AYA2016-80889-P, PID2019-108995GB-C21), the Consejer\'{\i}a de Econom\'{\i}a, Conocimiento, Empresas y Universidad of the Junta de Andaluc\'{\i}a (grant P18-FR-1769), the Consejo Superior de Investigaciones Cient\'{\i}ficas (grant 2019AEP112), and the State Agency for Research of the Spanish MCIU through the “Center of Excellence Severo Ochoa” award for the Instituto de Astrofísica de Andalucía (SEV-2017-0709).

\section*{Data Availability}
Most of the observational data used in this paper are taken from already published papers and references are mentioned.
The 86 GHz PA values of OJ~287 and any other data calculated in this paper are shown in the plots and will be available on request.




\bibliographystyle{mnras}
\bibliography{mnras_template} 

\begin{thebibliography}{}
\makeatletter
\relax
\def\mn@urlcharsother{\let\do\@makeother \do\$\do\&\do\#\do\^\do\_\do\%\do\~}
\def\mn@doi{\begingroup\mn@urlcharsother \@ifnextchar [ {\mn@doi@}
  {\mn@doi@[]}}
\def\mn@doi@[#1]#2{\def\@tempa{#1}\ifx\@tempa\@empty \href
  {http://dx.doi.org/#2} {doi:#2}\else \href {http://dx.doi.org/#2} {#1}\fi
  \endgroup}
\def\mn@eprint#1#2{\mn@eprint@#1:#2::\@nil}
\def\mn@eprint@arXiv#1{\href {http://arxiv.org/abs/#1} {{\tt arXiv:#1}}}
\def\mn@eprint@dblp#1{\href {http://dblp.uni-trier.de/rec/bibtex/#1.xml}
  {dblp:#1}}
\def\mn@eprint@#1:#2:#3:#4\@nil{\def\@tempa {#1}\def\@tempb {#2}\def\@tempc
  {#3}\ifx \@tempc \@empty \let \@tempc \@tempb \let \@tempb \@tempa \fi \ifx
  \@tempb \@empty \def\@tempb {arXiv}\fi \@ifundefined
  {mn@eprint@\@tempb}{\@tempb:\@tempc}{\expandafter \expandafter \csname
  mn@eprint@\@tempb\endcsname \expandafter{\@tempc}}}

\bibitem[\protect\citeauthoryear{{Abraham}}{{Abraham}}{2000}]{abraham00}
{Abraham} Z.,  2000, \aap, \href
  {https://ui.adsabs.harvard.edu/abs/2000A&A...355..915A} {355, 915}

\bibitem[\protect\citeauthoryear{{Agudo}, {Marscher}, {Jorstad}, {G{\'o}mez},
  {Perucho}, {Piner}, {Rioja}  \& {Dodson}}{{Agudo} et~al.}{2012}]{agu12}
{Agudo} I.,  {Marscher} A.~P.,  {Jorstad} S.~G.,  {G{\'o}mez} J.~L.,  {Perucho}
  M.,  {Piner} B.~G.,  {Rioja} M.,   {Dodson} R.,  2012, \mn@doi [\apj]
  {10.1088/0004-637X/747/1/63}, \href
  {https://ui.adsabs.harvard.edu/abs/2012ApJ...747...63A} {747, 63}

\bibitem[\protect\citeauthoryear{{Barker} \& {O'Connell}}{{Barker} \&
  {O'Connell}}{1979}]{BC79}
{Barker} B.~M.,  {O'Connell} R.~F.,  1979, \mn@doi [General Relativity and
  Gravitation] {10.1007/BF00756587}, \href
  {https://ui.adsabs.harvard.edu/abs/1979GReGr..11..149B} {11, 149}

\bibitem[\protect\citeauthoryear{{Blanchet}}{{Blanchet}}{2014}]{bla14}
{Blanchet} L.,  2014, \mn@doi [Living Reviews in Relativity]
  {10.12942/lrr-2014-2}, \href
  {http://adsabs.harvard.edu/abs/2014LRR....17....2B} {17, 2}

\bibitem[\protect\citeauthoryear{{Blandford} \& {Payne}}{{Blandford} \&
  {Payne}}{1982}]{BP82}
{Blandford} R.~D.,  {Payne} D.~G.,  1982, \mn@doi [\mnras]
  {10.1093/mnras/199.4.883}, \href
  {https://ui.adsabs.harvard.edu/abs/1982MNRAS.199..883B} {199, 883}

\bibitem[\protect\citeauthoryear{{Blandford} \& {Znajek}}{{Blandford} \&
  {Znajek}}{1977}]{BZ77}
{Blandford} R.~D.,  {Znajek} R.~L.,  1977, \mn@doi [\mnras]
  {10.1093/mnras/179.3.433}, \href
  {https://ui.adsabs.harvard.edu/abs/1977MNRAS.179..433B} {179, 433}

\bibitem[\protect\citeauthoryear{{Britzen} et~al.,}{{Britzen}
  et~al.}{2018}]{bri18}
{Britzen} S.,  et~al., 2018, \mn@doi [\mnras] {10.1093/mnras/sty1026}, \href
  {https://ui.adsabs.harvard.edu/abs/2018MNRAS.478.3199B} {478, 3199}

\bibitem[\protect\citeauthoryear{{Burrows} et~al.,}{{Burrows}
  et~al.}{2011}]{burrows2011}
{Burrows} D.~N.,  et~al., 2011, \mn@doi [\nat] {10.1038/nature10374}, \href
  {https://ui.adsabs.harvard.edu/abs/2011Natur.476..421B} {476, 421}

\bibitem[\protect\citeauthoryear{{Caproni} \& {Abraham}}{{Caproni} \&
  {Abraham}}{2004}]{caproni04}
{Caproni} A.,  {Abraham} Z.,  2004, in {Storchi-Bergmann} T.,  {Ho} L.~C.,
  {Schmitt} H.~R.,  eds,  IAU Symposium Vol. 222, The Interplay Among Black
  Holes, Stars and ISM in Galactic Nuclei. pp 83--84,
  \mn@doi{10.1017/S1743921304001541}

\bibitem[\protect\citeauthoryear{{Cohen}}{{Cohen}}{2017}]{coh17}
{Cohen} M.,  2017, \mn@doi [Galaxies] {10.3390/galaxies5010012}, \href
  {https://ui.adsabs.harvard.edu/abs/2017Galax...5...12C} {5, 12}

\bibitem[\protect\citeauthoryear{{Dey} et~al.,}{{Dey} et~al.}{2018}]{dey18}
{Dey} L.,  et~al., 2018, \mn@doi [\apj] {10.3847/1538-4357/aadd95}, \href
  {https://ui.adsabs.harvard.edu/abs/2018ApJ...866...11D} {866, 11}

\bibitem[\protect\citeauthoryear{{Dey} et~al.,}{{Dey} et~al.}{2019}]{dey19a}
{Dey} L.,  et~al., 2019, \mn@doi [Universe] {10.3390/universe5050108}, \href
  {https://ui.adsabs.harvard.edu/abs/2019Univ....5..108D} {5, 108}

\bibitem[\protect\citeauthoryear{{Event Horizon Telescope Collaboration}
  et~al.,}{{Event Horizon Telescope Collaboration} et~al.}{2019}]{EHT19a}
{Event Horizon Telescope Collaboration} et~al., 2019, \mn@doi [\apjl]
  {10.3847/2041-8213/ab0ec7}, \href
  {http://adsabs.harvard.edu/abs/2019ApJ...875L...1E} {875, L1}

\bibitem[\protect\citeauthoryear{{Feroz}, {Hobson}  \& {Bridges}}{{Feroz}
  et~al.}{2009}]{Feroz2009}
{Feroz} F.,  {Hobson} M.~P.,   {Bridges} M.,  2009, \mn@doi [\mnras]
  {10.1111/j.1365-2966.2009.14548.x}, \href
  {https://ui.adsabs.harvard.edu/abs/2009MNRAS.398.1601F} {398, 1601}

\bibitem[\protect\citeauthoryear{{Fossati}, {Maraschi}, {Celotti}, {Comastri}
  \& {Ghisellini}}{{Fossati} et~al.}{1998}]{GG98}
{Fossati} G.,  {Maraschi} L.,  {Celotti} A.,  {Comastri} A.,   {Ghisellini} G.,
   1998, \mn@doi [\mnras] {10.1046/j.1365-8711.1998.01828.x}, \href
  {https://ui.adsabs.harvard.edu/abs/1998MNRAS.299..433F} {299, 433}

\bibitem[\protect\citeauthoryear{{Gabuzda}, {Wardle}  \& {Roberts}}{{Gabuzda}
  et~al.}{1989}]{gabuzda89}
{Gabuzda} D.~C.,  {Wardle} J. F.~C.,   {Roberts} D.~H.,  1989, \mn@doi [\apjl]
  {10.1086/185361}, \href
  {https://ui.adsabs.harvard.edu/abs/1989ApJ...336L..59G} {336, L59}

\bibitem[\protect\citeauthoryear{{Giannios} \& {Metzger}}{{Giannios} \&
  {Metzger}}{2011}]{giannios2011}
{Giannios} D.,  {Metzger} B.~D.,  2011, \mn@doi [\mnras]
  {10.1111/j.1365-2966.2011.19188.x}, \href
  {https://ui.adsabs.harvard.edu/abs/2011MNRAS.416.2102G} {416, 2102}

\bibitem[\protect\citeauthoryear{{Hawley} \& {Krolik}}{{Hawley} \&
  {Krolik}}{2001}]{HK01}
{Hawley} J.~F.,  {Krolik} J.~H.,  2001, \mn@doi [\apj] {10.1086/318678}, \href
  {https://ui.adsabs.harvard.edu/abs/2001ApJ...548..348H} {548, 348}

\bibitem[\protect\citeauthoryear{{Hodgson} et~al.,}{{Hodgson}
  et~al.}{2017}]{hodgson17}
{Hodgson} J.~A.,  et~al., 2017, \mn@doi [\aap] {10.1051/0004-6361/201526727},
  \href {https://ui.adsabs.harvard.edu/abs/2017A&A...597A..80H} {597, A80}

\bibitem[\protect\citeauthoryear{{Hogg}, {Bovy}  \& {Lang}}{{Hogg}
  et~al.}{2010}]{Hogg2010}
{Hogg} D.~W.,  {Bovy} J.,   {Lang} D.,  2010, arXiv e-prints, \href
  {https://ui.adsabs.harvard.edu/abs/2010arXiv1008.4686H} {p. arXiv:1008.4686}

\bibitem[\protect\citeauthoryear{{Kembhavi} \& {Narlikar}}{{Kembhavi} \&
  {Narlikar}}{1999}]{KN99}
{Kembhavi} A.~K.,  {Narlikar} J.~V.,  1999, {Quasars and active galactic nuclei
  : an introduction}.
Cambridge University Press, \mn@doi{10.1017/CBO9781139174404}

\bibitem[\protect\citeauthoryear{{King}, {Pringle}  \& {Livio}}{{King}
  et~al.}{2007}]{king07}
{King} A.~R.,  {Pringle} J.~E.,   {Livio} M.,  2007, \mn@doi [\mnras]
  {10.1111/j.1365-2966.2007.11556.x}, \href
  {https://ui.adsabs.harvard.edu/abs/2007MNRAS.376.1740K} {376, 1740}

\bibitem[\protect\citeauthoryear{{Komossa} \& {Zensus}}{{Komossa} \&
  {Zensus}}{2016}]{komossa2016}
{Komossa} S.,  {Zensus} J.~A.,  2016, in {Meiron} Y.,  {Li} S.,  {Liu} F.~K.,
  {Spurzem} R.,  eds,  IAU Symposium Vol. 312, Star Clusters and Black Holes in
  Galaxies across Cosmic Time. pp 13--25 (\mn@eprint {arXiv} {1502.05720}),
  \mn@doi{10.1017/S1743921315007395}

\bibitem[\protect\citeauthoryear{{Komossa}, {Grupe}, {Parker}, {Valtonen},
  {G{\'o}mez}, {Gopakumar}  \& {Dey}}{{Komossa} et~al.}{2020}]{komossa2020}
{Komossa} S.,  {Grupe} D.,  {Parker} M.~L.,  {Valtonen} M.~J.,  {G{\'o}mez}
  J.~L.,  {Gopakumar} A.,   {Dey} L.,  2020, \mn@doi [\mnras]
  {10.1093/mnrasl/slaa125}, \href
  {https://ui.adsabs.harvard.edu/abs/2020MNRAS.498L..35K} {498, L35}

\bibitem[\protect\citeauthoryear{K\"onigsd\"orffer \&
  Gopakumar}{K\"onigsd\"orffer \& Gopakumar}{2005}]{KG05}
K\"onigsd\"orffer C.,  Gopakumar A.,  2005, \mn@doi [Phys. Rev. D]
  {10.1103/PhysRevD.71.024039}, 71, 024039

\bibitem[\protect\citeauthoryear{{Laine} et~al.,}{{Laine}
  et~al.}{2020}]{laine20}
{Laine} S.,  et~al., 2020, \mn@doi [\apjl] {10.3847/2041-8213/ab79a4}, \href
  {https://ui.adsabs.harvard.edu/abs/2020ApJ...894L...1L} {894, L1}

\bibitem[\protect\citeauthoryear{{Lehto} \& {Valtonen}}{{Lehto} \&
  {Valtonen}}{1996}]{LV96}
{Lehto} H.~J.,  {Valtonen} M.~J.,  1996, \mn@doi [\apj] {10.1086/176962}, \href
  {https://ui.adsabs.harvard.edu/abs/1996ApJ...460..207L} {460, 207}

\bibitem[\protect\citeauthoryear{{Lico} et~al.,}{{Lico}
  et~al.}{2020}]{Lico2020}
{Lico} R.,  et~al., 2020, \mn@doi [\aap] {10.1051/0004-6361/201936564}, \href
  {https://ui.adsabs.harvard.edu/abs/2020A&A...634A..87L} {634, A87}

\bibitem[\protect\citeauthoryear{{Lister}, {Aller}, {Aller}, {Hodge}, {Homan},
  {Kovalev}, {Pushkarev}  \& {Savolainen}}{{Lister} et~al.}{2018}]{mojave2018}
{Lister} M.~L.,  {Aller} M.~F.,  {Aller} H.~D.,  {Hodge} M.~A.,  {Homan} D.~C.,
   {Kovalev} Y.~Y.,  {Pushkarev} A.~B.,   {Savolainen} T.,  2018, VizieR Online
  Data Catalog, \href {https://ui.adsabs.harvard.edu/abs/2018yCat..22340012L}
  {p. J/ApJS/234/12}

\bibitem[\protect\citeauthoryear{{Lynden-Bell}}{{Lynden-Bell}}{1969}]{lyn69}
{Lynden-Bell} D.,  1969, \mn@doi [\nat] {10.1038/223690a0}, \href
  {http://adsabs.harvard.edu/abs/1969Natur.223..690L} {223, 690}

\bibitem[\protect\citeauthoryear{{Marscher}, {Jorstad}, {G{\'o}mez}, {Aller},
  {Ter{\"a}sranta}, {Lister}  \& {Stirling}}{{Marscher}
  et~al.}{2002}]{Marscher02}
{Marscher} A.~P.,  {Jorstad} S.~G.,  {G{\'o}mez} J.-L.,  {Aller} M.~F.,
  {Ter{\"a}sranta} H.,  {Lister} M.~L.,   {Stirling} A.~M.,  2002, \mn@doi
  [\nat] {10.1038/nature00772}, \href
  {https://ui.adsabs.harvard.edu/abs/2002Natur.417..625M} {417, 625}

\bibitem[\protect\citeauthoryear{{Marscher} et~al.,}{{Marscher}
  et~al.}{2018}]{marscher2018}
{Marscher} A.~P.,  et~al., 2018, \mn@doi [\apj] {10.3847/1538-4357/aae4de},
  \href {https://ui.adsabs.harvard.edu/abs/2018ApJ...867..128M} {867, 128}

\bibitem[\protect\citeauthoryear{{Miller}}{{Miller}}{1976}]{mil76}
{Miller} R.~H.,  1976, \mn@doi [Journal of Computational Physics]
  {10.1016/0021-9991(76)90038-3}, \href
  {https://ui.adsabs.harvard.edu/abs/1976JCoPh..21..400M} {21, 400}

\bibitem[\protect\citeauthoryear{{Pihajoki} et~al.,}{{Pihajoki}
  et~al.}{2013}]{pih13}
{Pihajoki} P.,  et~al., 2013, \mn@doi [\apj] {10.1088/0004-637X/764/1/5}, \href
  {https://ui.adsabs.harvard.edu/abs/2013ApJ...764....5P} {764, 5}

\bibitem[\protect\citeauthoryear{{Pushkarev}, {Hovatta}, {Kovalev}, {Lister},
  {Lobanov}, {Savolainen}  \& {Zensus}}{{Pushkarev}
  et~al.}{2012}]{pushkarev2012}
{Pushkarev} A.~B.,  {Hovatta} T.,  {Kovalev} Y.~Y.,  {Lister} M.~L.,  {Lobanov}
  A.~P.,  {Savolainen} T.,   {Zensus} J.~A.,  2012, \mn@doi [\aap]
  {10.1051/0004-6361/201219173}, \href
  {https://ui.adsabs.harvard.edu/abs/2012A&A...545A.113P} {545, A113}

\bibitem[\protect\citeauthoryear{{Pushkarev}, {Kovalev}, {Lister}  \&
  {Savolainen}}{{Pushkarev} et~al.}{2017}]{Pushkarev2017}
{Pushkarev} A.~B.,  {Kovalev} Y.~Y.,  {Lister} M.~L.,   {Savolainen} T.,  2017,
  \mn@doi [\mnras] {10.1093/mnras/stx854}, \href
  {https://ui.adsabs.harvard.edu/abs/2017MNRAS.468.4992P} {468, 4992}

\bibitem[\protect\citeauthoryear{{Qian}}{{Qian}}{2018}]{qian18}
{Qian} S.,  2018, arXiv e-prints, \href
  {https://ui.adsabs.harvard.edu/abs/2018arXiv181111514Q} {p. arXiv:1811.11514}

\bibitem[\protect\citeauthoryear{{Roberts}, {Gabuzda}  \& {Wardle}}{{Roberts}
  et~al.}{1987}]{roberts87}
{Roberts} D.~H.,  {Gabuzda} D.~C.,   {Wardle} J. F.~C.,  1987, \mn@doi [\apj]
  {10.1086/165849}, \href
  {https://ui.adsabs.harvard.edu/abs/1987ApJ...323..536R} {323, 536}

\bibitem[\protect\citeauthoryear{{Sakimoto} \& {Coroniti}}{{Sakimoto} \&
  {Coroniti}}{1981}]{SC81}
{Sakimoto} P.~J.,  {Coroniti} F.~V.,  1981, \mn@doi [\apj] {10.1086/159005},
  \href {https://ui.adsabs.harvard.edu/abs/1981ApJ...247...19S} {247, 19}

\bibitem[\protect\citeauthoryear{{Shakura} \& {Sunyaev}}{{Shakura} \&
  {Sunyaev}}{1973}]{SS73}
{Shakura} N.~I.,  {Sunyaev} R.~A.,  1973, \aap, \href
  {https://ui.adsabs.harvard.edu/abs/1973A&A....24..337S} {500, 33}

\bibitem[\protect\citeauthoryear{{Sillanpaa}, {Haarala}, {Valtonen},
  {Sundelius}  \& {Byrd}}{{Sillanpaa} et~al.}{1988}]{sil1988}
{Sillanpaa} A.,  {Haarala} S.,  {Valtonen} M.~J.,  {Sundelius} B.,   {Byrd}
  G.~G.,  1988, \mn@doi [\apj] {10.1086/166033}, \href
  {https://ui.adsabs.harvard.edu/abs/1988ApJ...325..628S} {325, 628}

\bibitem[\protect\citeauthoryear{{Skilling}}{{Skilling}}{2004}]{Skilling2004}
{Skilling} J.,  2004, in {Fischer} R.,  {Preuss} R.,   {Toussaint} U.~V.,  eds,
   American Institute of Physics Conference Series Vol. 735, American Institute
  of Physics Conference Series. pp 395--405, \mn@doi{10.1063/1.1835238}

\bibitem[\protect\citeauthoryear{{Sundelius}, {Wahde}, {Lehto}  \&
  {Valtonen}}{{Sundelius} et~al.}{1997}]{sun97}
{Sundelius} B.,  {Wahde} M.,  {Lehto} H.~J.,   {Valtonen} M.~J.,  1997, \mn@doi
  [\apj] {10.1086/304331}, \href
  {https://ui.adsabs.harvard.edu/abs/1997ApJ...484..180S} {484, 180}

\bibitem[\protect\citeauthoryear{{Tateyama}}{{Tateyama}}{2013}]{tat13}
{Tateyama} C.~E.,  2013, \mn@doi [\apjs] {10.1088/0067-0049/205/2/15}, \href
  {https://ui.adsabs.harvard.edu/abs/2013ApJS..205...15T} {205, 15}

\bibitem[\protect\citeauthoryear{{Tateyama} \& {Kingham}}{{Tateyama} \&
  {Kingham}}{2004}]{tateyama04}
{Tateyama} C.~E.,  {Kingham} K.~A.,  2004, \mn@doi [\apj] {10.1086/392524},
  \href {https://ui.adsabs.harvard.edu/abs/2004ApJ...608..149T} {608, 149}

\bibitem[\protect\citeauthoryear{{Tateyama}, {Kingham}, {Kaufmann}, {Piner},
  {Botti}  \& {de Lucena}}{{Tateyama} et~al.}{1999}]{tateyama99}
{Tateyama} C.~E.,  {Kingham} K.~A.,  {Kaufmann} P.,  {Piner} B.~G.,  {Botti}
  L.~C.~L.,   {de Lucena} A.~M.~P.,  1999, \mn@doi [\apj] {10.1086/307475},
  \href {https://ui.adsabs.harvard.edu/abs/1999ApJ...520..627T} {520, 627}

\bibitem[\protect\citeauthoryear{{Tchekhovskoy}, {Metzger}, {Giannios}  \&
  {Kelley}}{{Tchekhovskoy} et~al.}{2014}]{tchekhovskoy2014}
{Tchekhovskoy} A.,  {Metzger} B.~D.,  {Giannios} D.,   {Kelley} L.~Z.,  2014,
  \mn@doi [\mnras] {10.1093/mnras/stt2085}, \href
  {https://ui.adsabs.harvard.edu/abs/2014MNRAS.437.2744T} {437, 2744}

\bibitem[\protect\citeauthoryear{{Urry} \& {Padovani}}{{Urry} \&
  {Padovani}}{1995}]{Urry95}
{Urry} C.~M.,  {Padovani} P.,  1995, \mn@doi [\pasp] {10.1086/133630}, \href
  {https://ui.adsabs.harvard.edu/abs/1995PASP..107..803U} {107, 803}

\bibitem[\protect\citeauthoryear{{Valtonen}}{{Valtonen}}{2007}]{val07}
{Valtonen} M.~J.,  2007, \mn@doi [\apj] {10.1086/512801}, \href
  {https://ui.adsabs.harvard.edu/abs/2007ApJ...659.1074V} {659, 1074}

\bibitem[\protect\citeauthoryear{{Valtonen} \& {Pihajoki}}{{Valtonen} \&
  {Pihajoki}}{2013}]{VP13}
{Valtonen} M.,  {Pihajoki} P.,  2013, \mn@doi [\aap]
  {10.1051/0004-6361/201321754}, \href
  {https://ui.adsabs.harvard.edu/abs/2013A&A...557A..28V} {557, A28}

\bibitem[\protect\citeauthoryear{{Valtonen} \& {Wiik}}{{Valtonen} \&
  {Wiik}}{2012}]{VW12}
{Valtonen} M.~J.,  {Wiik} K.,  2012, \mn@doi [\mnras]
  {10.1111/j.1365-2966.2011.20009.x}, \href
  {https://ui.adsabs.harvard.edu/abs/2012MNRAS.421.1861V} {421, 1861}

\bibitem[\protect\citeauthoryear{{Valtonen} et~al.,}{{Valtonen}
  et~al.}{2006}]{val06b}
{Valtonen} M.~J.,  et~al., 2006, \mn@doi [\apj] {10.1086/504884}, \href
  {https://ui.adsabs.harvard.edu/abs/2006ApJ...646...36V} {646, 36}

\bibitem[\protect\citeauthoryear{{Valtonen} et~al.,}{{Valtonen}
  et~al.}{2008}]{val08}
{Valtonen} M.~J.,  et~al., 2008, \mn@doi [\nat] {10.1038/nature06896}, \href
  {https://ui.adsabs.harvard.edu/abs/2008Natur.452..851V} {452, 851}

\bibitem[\protect\citeauthoryear{{Valtonen} et~al.,}{{Valtonen}
  et~al.}{2010}]{val10}
{Valtonen} M.~J.,  et~al., 2010, \mn@doi [\apj] {10.1088/0004-637X/709/2/725},
  \href {https://ui.adsabs.harvard.edu/abs/2010ApJ...709..725V} {709, 725}

\bibitem[\protect\citeauthoryear{{Valtonen}, {Lehto}, {Takalo}  \&
  {Sillanp{\"a}{\"a}}}{{Valtonen} et~al.}{2011a}]{val11a}
{Valtonen} M.~J.,  {Lehto} H.~J.,  {Takalo} L.~O.,   {Sillanp{\"a}{\"a}} A.,
  2011a, \mn@doi [\apj] {10.1088/0004-637X/729/1/33}, \href
  {https://ui.adsabs.harvard.edu/abs/2011ApJ...729...33V} {729, 33}

\bibitem[\protect\citeauthoryear{{Valtonen}, {Mikkola}, {Lehto}, {Gopakumar},
  {Hudec}  \& {Polednikova}}{{Valtonen} et~al.}{2011b}]{val11b}
{Valtonen} M.~J.,  {Mikkola} S.,  {Lehto} H.~J.,  {Gopakumar} A.,  {Hudec} R.,
   {Polednikova} J.,  2011b, \mn@doi [\apj] {10.1088/0004-637X/742/1/22}, \href
  {https://ui.adsabs.harvard.edu/abs/2011ApJ...742...22V} {742, 22}

\bibitem[\protect\citeauthoryear{{Valtonen} et~al.,}{{Valtonen}
  et~al.}{2016}]{val16}
{Valtonen} M.~J.,  et~al., 2016, \mn@doi [\apjl] {10.3847/2041-8205/819/2/L37},
  \href {https://ui.adsabs.harvard.edu/abs/2016ApJ...819L..37V} {819, L37}

\bibitem[\protect\citeauthoryear{{Valtonen} et~al.,}{{Valtonen}
  et~al.}{2019}]{val19}
{Valtonen} M.~J.,  et~al., 2019, \mn@doi [\apj] {10.3847/1538-4357/ab3573},
  \href {https://ui.adsabs.harvard.edu/abs/2019ApJ...882...88V} {882, 88}

\bibitem[\protect\citeauthoryear{{Will} \& {Maitra}}{{Will} \&
  {Maitra}}{2017}]{WM17}
{Will} C.~M.,  {Maitra} M.,  2017, \mn@doi [\prd] {10.1103/PhysRevD.95.064003},
  \href {https://ui.adsabs.harvard.edu/abs/2017PhRvD..95f4003W} {95, 064003}

\bibitem[\protect\citeauthoryear{{Worrall} et~al.,}{{Worrall}
  et~al.}{1982}]{worrall82}
{Worrall} D.~M.,  et~al., 1982, \mn@doi [\apj] {10.1086/160352}, \href
  {https://ui.adsabs.harvard.edu/abs/1982ApJ...261..403W} {261, 403}

\bibitem[\protect\citeauthoryear{{Yang}, {Paragi}, {van der Horst}, {Gurvits},
  {Campbell}, {Giannios}, {An}  \& {Komossa}}{{Yang} et~al.}{2016}]{yang2016}
{Yang} J.,  {Paragi} Z.,  {van der Horst} A.~J.,  {Gurvits} L.~I.,  {Campbell}
  R.~M.,  {Giannios} D.,  {An} T.,   {Komossa} S.,  2016, \mn@doi [\mnras]
  {10.1093/mnrasl/slw107}, \href
  {https://ui.adsabs.harvard.edu/abs/2016MNRAS.462L..66Y} {462, L66}

\bibitem[\protect\citeauthoryear{{Zauderer} et~al.,}{{Zauderer}
  et~al.}{2011}]{zauderer2011}
{Zauderer} B.~A.,  et~al., 2011, \mn@doi [\nat] {10.1038/nature10366}, \href
  {https://ui.adsabs.harvard.edu/abs/2011Natur.476..425Z} {476, 425}

\bibitem[\protect\citeauthoryear{{Zauderer}, {Berger}, {Margutti}, {Pooley},
  {Sari}, {Soderberg}, {Brunthaler}  \& {Bietenholz}}{{Zauderer}
  et~al.}{2013}]{zauderer2013}
{Zauderer} B.~A.,  {Berger} E.,  {Margutti} R.,  {Pooley} G.~G.,  {Sari} R.,
  {Soderberg} A.~M.,  {Brunthaler} A.,   {Bietenholz} M.~F.,  2013, \mn@doi
  [\apj] {10.1088/0004-637X/767/2/152}, \href
  {https://ui.adsabs.harvard.edu/abs/2013ApJ...767..152Z} {767, 152}

\makeatother
\end{thebibliography}








\bsp{}	
\label{lastpage}
\end{document}